\documentclass[12pt]{article}

\usepackage{amsmath}
\usepackage{amssymb}
\usepackage{amsfonts}
\usepackage{latexsym}

\catcode `\@=11 \@addtoreset{equation}{section}
\def\theequation{\arabic{section}.\arabic{equation}}
\catcode `\@=12

\newcommand{\be}{\begin{equation}}
\newcommand{\en}{\end{equation}}
\newcommand{\bea}{\begin{eqnarray}}
\newcommand{\ena}{\end{eqnarray}}
\newcommand{\beano}{\begin{eqnarray*}}
\newcommand{\enano}{\end{eqnarray*}}
\newcommand{\bee}{\begin{enumerate}}
\newcommand{\ene}{\end{enumerate}}

\newcommand{\mb}{\mathbb}
\newcommand{\A}{{\mathfrak A}}
\newcommand{\B}{{\mathfrak B}}
\newcommand{\Ao}{{\mathfrak A}_0}

\newcommand{\R}{\mathbb{R}}
\newcommand{\mc}{\mathcal}
\newcommand{\norm}[1]{ \parallel #1 \parallel}
\newcommand{\D}{{\mc D}}
\newcommand{\LL}{\mc L}

\newcommand{\id}{{\mb I}}

\newcommand{\Sys}{{\cal S}}
\newcommand{\E}{{\cal E}}
\newcommand{\F}{{\cal F}}
\newcommand{\I}{{\mathbb{I}}}

\newcommand{\Lc}{{\cal L}}
\newcommand{\C}{{\cal C}}
\newcommand{\1}{1 \!\! 1}
\newcommand{\N}{\mc N}

\newcommand{\LD}{{\LL}^\dagger (\D)}
\newcommand{\LDD}{{\LL} (\D_\pi,\D_\pi')}

\newcommand{\restr}{\upharpoonright}
\newcommand{\Hil}{\mc H}

\newtheorem{thm}{Theorem}

\newtheorem{lemma}[thm]{Lemma}
\newtheorem{prop}[thm]{Proposition}
\newtheorem{defn}[thm]{Definition}

\catcode `\@=11 \@addtoreset{equation}{section}
\catcode `\@=12

\begin{document}

\thispagestyle{empty}

\vspace*{1cm}

\begin{center}
{\Large \bf Algebras of unbounded operators and physical
applications:
a survey}   \vspace{2cm}\\

{\large F. Bagarello}
\vspace{3mm}\\
  Dipartimento di Metodi e Modelli Matematici,
Facolt\`a di Ingegneria, Universit\`a di Palermo, \\Viale delle Scienze, I-90128  Palermo, Italy\\
e-mail: bagarell@unipa.it\\home page: www.unipa.it/$^\sim$bagarell
\vspace{4mm}\\

\end{center}

\vspace*{2cm}

\begin{abstract}
\noindent After an historical introduction on the standard algebraic
approach to quantum mechanics of large systems we review the basic
mathematical aspects of the algebras of unbounded operators. After
that we discuss in some details their relevance in physical
applications.
\end{abstract}

\newpage
\section{Introduction}

During the past 20 years a long series of papers concerning algebras
of unbounded operators appeared in the literature, papers which,
though being originally motivated by physical arguments, contain
essentially no physics at all. On the contrary the mathematical
aspects of these algebras have been analyzed in many details and
this analysis produced, up to now, the monographs \cite{schu} and
\cite{aitbook}. Some physics appeared first in \cite{lass} and
\cite{timm}, in the attempt to describe systems with a {\it very
large} ($10^{24}$) number of degrees of freedom, following some
general ideas originally proposed in the famous paper of Haag and
Kastler, \cite{hk}.

These authors consider, as widely discussed in the literature,
\cite{br},  systems with {\it infinite } degrees of freedom because,
in this way, a simpler approach to, e.g., phase transitions and
collective phenomena can be settled up. However, moving from a {\it
large but finite} to an {\it infinite} number of degrees of freedom
one has to build up a mathematical apparatus which is rather
sophisticated and, as we will see, not yet completely fixed.

More recently other physical applications of algebras of unbounded
operators have been proposed by the present author and others, see
\cite{bag1,bag2,bt1,bt2,bt3,bt4,bit1,bit2} for instance. In our
opinion it is time to review  some of these results, trying to
connect as much as possible these with the {\it original } results
based on C*-algebras.

The paper, which is meant to be very pedagogical, is organized as
follows: in Section II we give an introduction to
 non relativistic ordinary quantum mechanics (i.e. quantum mechanics for systems with a finite
 number of degrees of freedom), useful to fix the notation and some preliminary
 ideas. Section III is devoted to
 a longer review to non relativistic quantum mechanics for systems with infinite degrees of
 freedom, with a particular interest for
some physically relevant results and  for open problems. In Section
IV we introduce some mathematical definitions and results concerning
algebras of unbounded operators, while their physical applications
are given in Section V. Our conclusions and our future projects are
finally contained in Section VI. To keep the paper self-contained we
have also added two Appendices. In the first one we give the general
construction of the algebraic settings which extends the Haag and
Kastler's construction, while in the second appendix we give a list
of information of functional analysis which may be useful to some
non particularly mathematically minded.

\section{Ordinary (non relativistic) quantum mechanics}

This and the next sections are heavily based on
\cite{sewbook1,sewbook2}, to which we refer for further details.

The usual description of non relativistic quantum mechanics, as it
is taught in  many textbooks, is given in some fixed Hilbert $\Hil$
space as follows:

each observable $A$ of the physical system corresponds to a
self-adjoint operator $\hat A$ in $\Hil$;

the pure states of the physical system corresponds to normalized
vectors of $\Hil$;

the expectation values of $A$ correspond to the following mean
values: $<\psi,\hat A\psi>=\rho_\psi(\hat A)=tr(P_\psi\,\hat A)$,
where we have also introduced a projector operator $P_\psi$ on
$\psi$ and $tr$ is the trace in $\Hil$;

the states which are not pure, i.e. the mixed states, correspond to
convex linear combinations $\hat\rho=\sum_j\,w_j\,\rho_{\psi_n}$,
with $\sum_j\,w_j=1$ and $w_j\geq 0$ for all $j$;

the dynamics (in the Schr\"odinger representation) is given by a
unitary operator
 $U_t:=e^{i H t/\hbar}$, where $H$ is the self-adjoint energy operator, as follows: $\hat\rho
\rightarrow \hat\rho_t=U_t^*\hat\rho U_t$. In the Heisenberg
representation the states do not evolve in time while the operators
do, following the  {\it dual} rule:
 $\hat A \rightarrow \hat A_t=U_t\hat A U_t^*$, and the Heisenberg
 equation of motion is satisfied:
$\frac{d}{dt}\hat A_t=\frac{i}{\hbar}[H,\hat A_t]$. It is very well
known that these two different strategies have the same physical
content: $\hat\rho(\hat A_t)=\hat\rho_t(\hat A).$

A different description of a quantum mechanical system, which is
more useful for its extensions to quantum systems with infinite
degrees of freedom, is the {\it algebraic} description.

In this approach the observables are elements of a C*-algebra $\A$
(which coincides with  $B(\Hil)$ for some Hilbert space $\Hil$).
This means, first of all, that  $\A$ is a vector space over
$\mathbb{C}$ with a multiplication law such that $\forall A,B\in\A$,
$AB\in\A$. Also, two such elements can be summed up and the
following properties hold: $\forall A,B,C\in\A$ and $\forall
\alpha,\beta\in\mathbb{C}$ we have
$$
A(BC)=(AB)C,\hspace{5mm} A(B+C)=AB+AC, \hspace{5mm} (\alpha A)(\beta
B)=\alpha\beta (AB).
$$
An involution is a map $*:\A\rightarrow \A$ such that
$$
A^{**}=A,\hspace{5mm} (AB)^*=B^*A^*,\hspace{5mm} (\alpha A+\beta
B)^*=\overline{\alpha}\,A^*+\overline{\beta}\,B^*
$$
A *-algebra $\A$ is an algebra with an involution *. $\A$ is  a
{\it normed algebra} if there exists a map, {\it the norm of the
algebra}, $\|.\|:\A\rightarrow \mathbb{R}_+$, such that:
$$
\|A\|\geq 0,\hspace{5mm} \|A\|=0 \Longleftrightarrow
A=0,\hspace{5mm} \|\alpha A\|=|\alpha|\,\|A\|,$$ $$\|A+B\|\leq
\|A\|+\|B\|,\hspace{5mm}\|AB\|\leq \|A\|\,\|B\|.
$$

If $\A$ is complete wrt $\|.\|$ then it is called a {\it   Banach
algebra}, or a {\it   Banach *-algebra} if $\|A^*\|=\|A\|$.

Finally, a {\it   C*-algebra} is a Banach *-algebra with the
property $\|A^*A\|=\|A\|^2$.

\vspace{2mm}

{\bf   Remarks:--}(1) Using this description of our physical system
we pay more attention on the {\em rules between the elements which
describe the system} rather than {\em on the way in which these
elements concretely act on a given Hilbert space}.

(2) All the  C*-algebras are isomorphic to a norm-closed,
*-closed, algebra of bounded operators on a certain Hilbert space.

(3) All the  abelian C*-algebras are isomorphic to the *-algebra of
continuous functions, over a locally compact Hausdorff space $X$,
which vanish at infinity, $C_o(X)$.

\vspace{3mm}

 The {\it states} are linear, positive and normalized
functional on $\A$, which looks like $\rho(\hat A)=tr(\hat\rho A)$,
when $\A=B(\Hil)$,   $\hat\rho$ is a trace-class operator and $tr$
is the trace on $\Hil$. This means in particular that
$$
\rho(\alpha_1 A+\alpha_2 B)=\alpha_1\rho(A)+\alpha_2\rho(B)$$ and
that
$$
\rho(A^*A)\geq 0;\hspace{5mm} \rho(\1)=1.
$$
An immediate consequence of these assumptions, and in particular of
the positivity of $\rho$, is that $\rho$ is also continuous, i.e.
that $|\rho(A)|\leq \|A\|$ for all $A\in \A$.

The dynamics in the Heisenberg representation for conservative
quantum systems, i.e. for systems which do not interact with the
environment, is given by the map

$$\A\ni A\rightarrow \alpha^t(A)=U_tAU_t^*\in \A,\,\forall
t$$ which defines a 1-parameter group of *-automorphisms of $\A$
satisfying the following conditions
$$\alpha^t(\lambda A)=\lambda \alpha^t(A),\hspace{5mm}
\alpha^t(A+B)=\alpha^t(A)+\alpha^t(B),$$
$$\alpha^t(AB)=\alpha^t(A)\,\alpha^t(B),\hspace{5mm}
\|\alpha^t(A)\|=\|A\|,\, \mbox{ and }\,
\alpha^{t+s}=\alpha^t\,\alpha^s.$$

{\bf Remark:--} in the Schr\"odinger representation the time
evolution is the dual of the one above, i.e. $\hat \rho\rightarrow
\hat\rho_t={\alpha^t}^*\hat\rho$.

\vspace{2mm}

The reason why this algebraic approach to ordinary quantum mechanics
is not very much used in the literature follows from the following
von Neumann uniqueness theorem: {\it for finite quantum mechanical
systems there exists only one irreducible representation} (but for
unitary equivalence):

let us consider, for instance, two operators $Q$ and $P$ such that
$[ Q, P]=i\hbar\mathbb{I}$. They can be irreducibly  represented on
$\Hil=\Lc^2(\mathbb{R})$ as follows: $ \hat q f(q)=qf(q)$, $ \hat
pf(q)=-i\hbar f'(q)$, $\forall f\in \Sys(\mathbb{R})$, which is
dense in $\Hil$. If now $\hat q', \hat p'$ is a (different)
irreducible representation of $ Q, P$ on a (different) Hilbert space
$\Hil'$, $[\hat q',\hat p']=i\hbar\mathbb{I}$,  then there exists an
unitary map $V:\Hil\rightarrow \Hil'$ such that $\hat q'=V\hat q
V^*$, $\hat p'=V\hat p V^*$. Notice that a more precise formulation
of von Neumann theorem should require the use of the Weyl unitary
operators, to avoid domain problems connected with the unboundedness
of $Q$ and $P$, \cite{sewbook2}.

This result can be interpreted as follows: there is no difference in
using an abstract C*-algebra or a given Hilbert space when dealing
with a quantum system with finite number of degrees of freedom since
two different (but irreducible) representations of the same algebra
are surely related by a unitary map and, for this reason, they are
physically equivalent. As we will discuss in the next section, this
is not what happens  in quantum mechanics for systems with infinite
degrees of freedom, $QM_\infty$ in the following, so that the two
descriptions became really different.

\section{A short review of (non relativistic) $QM_\infty$}

As just stated, when the degrees of freedom of a system increase
up to infinity, the uniqueness von Neumann theorem does not need
to hold, in the sense that the same physical system may have
several inequivalent representations.

A very simple example which exhibits such a feature is an infinite
spin chain, the so-called Ising model, whose (formal) hamiltonian is
$H=-J\sum_j \sigma_j^3\sigma_{j+1}^3$. Here $\sigma_j^3$ is the
third component of the Pauli matrices localized at the site $j$ of a
certain lattice. If $J>0$ the following vectors both minimize the
energy of the system:
$$\psi_0^{(+1)}=\ldots\otimes\uparrow\otimes\uparrow\otimes\uparrow\otimes\uparrow\ldots$$
 and
$$\psi_0^{(-1)}=\ldots\otimes\downarrow\otimes\downarrow\otimes\downarrow\otimes\downarrow\ldots$$
where $\uparrow$ and $\downarrow$ are eigenstates of $\sigma^3$ with
eigenvalues $+1$ and $-1$ respectively: $\sigma^3=\left(
\begin{array}{cc}
1 &  0  \\
0 & -1   \\
\end{array}
\right)$, $\uparrow=\left(\begin{array}{c}
1   \\
0    \\
\end{array}
\right)$ and $\downarrow=\left(\begin{array}{c}
0   \\
1    \\
\end{array}
\right)$. Furthermore, $\psi_0^{(\pm 1)}$ cannot be mapped into one
another by local actions, since for instance $\psi_0^{(+1)}$ can be
obtained from $\psi_0^{(-1)}$ only acting with $\sigma^1$ on {\it
each  site } of the infinite lattice! For this reason, there exists
no element $Y$ of a local algebra (see below) satisfying the
equality $Y\psi_0^{(+1)}=\psi_0^{(-1)}$. As we will show in a while,
these two states are related to two different representations of the
same abstract C*-algebra, representations which are  labeled by
different values of an order parameter, the so called {\it
magnetization}, $m\simeq<\psi_0^{(\pm 1)},\frac{1}{|V|}\,\sum_{j\in
V}\pi^{(\pm 1)}(\sigma_j^3)\,\psi_0^{(\pm 1)}>\,\rightarrow\pm 1$.
This implies, moreover, that the two representations, called here
$\pi^{(+1)}$ and $\pi^{(-1)}$, are necessarily {\it unitarely
inequivalent}, since they describe different physics.

It may be worth noticing also that this model exhibits a first
example of {\it spontaneous breaking of a symmetry}: the following
symmetry of the  $H$, $\gamma:\sigma_j^3\rightarrow -\sigma_j^3$, is
clearly not a symmetry of the ground state, meaning with this that
the two vectors $\psi_0^{(\pm 1)}$ are not left invariant by the map
$\gamma$. We will discuss in the following a consequence of this
property. \vspace{2mm}

\vspace{3mm}

This kind of physical systems can be properly discussed within the
framework of C*-algebras, as first proposed by Haag and Kastler,
\cite{hk}, whose construction goes as follows.

\vspace{1mm}

{\bf The algebra.} Let $\Sigma$ be a physical system with infinite
degrees of freedom, $V\subset \mathbb{R}^d$ a finite d-dimensional
region, $\Hil_V$ the related Hilbert space (whose construction
depends on $\Sigma$ and will be discussed in a moment),
$\A_V=B(\Hil_V)$ the associated C*-algebra of bounded operators
acting on $\Hil_V$ and let finally $H_V$ be the self-adjoint energy
operator for $\Sigma_V$, the restriction of $\Sigma$ in $V$.

The family  of algebras  $\{\A_V\}$ satisfies the following
properties:
\begin{itemize}
\item {\it isotony}: if $V_1\subset V_2$ then $\A_{V_1}\subset
\A_{V_2}$. Moreover $\|.\|_2\downharpoonright_{V_1}=\|.\|_1$
$(\Rightarrow \A_{V_1},\A_{V_2}\subset\A_{V_1\cup V_2})$;
\item if $V_1\cap V_2=\emptyset$ then
$[\A_{V_1},\A_{V_2}]=0$.
\end{itemize}
In particular this last property clearly shows the non relativistic
framework we are discussing here, since it simply means that two
operators localized in disjoint spatial regions are necessarily
independent, i.e. they must commute. Then we define {$
\A=\overline{\A_0}^{\|.\|},$} where
  {$\A_0=\cup_V \A_V$} and
$\A$ is the {\it quasi-local C*-algebra of the {\bf bounded}
observables.}

On this algebra  we can  introduce the   {\it spatial
translations} $\{\gamma_{ x}\}$, which is a group of
*-automorphisms of $\A$ satisfying the following: $\gamma_{
x}\A_V=\A_{V+ x}$, $\gamma_{ x_1}\gamma_{ x_2}= \gamma_{ x_1+
x_2}$.

It is now worth discussing briefly two examples of this
construction.

\vspace{2mm}

  {{\bf Example 1:} {\it discrete system}}

Let $X$ be an infinite lattice, $ 0\in X$, and $\Hil_{ 0}$ a finite
dimensional Hilbert space (e.g. $\Hil_{ 0}=\mathbb{C}^2$ for Pauli
matrices). Let $\Hil_{ x}$ a copy of $\Hil_{ 0}$ localized in the
lattice site $ x\in X$ and $\Hil_V=\otimes_{ x\in V} \Hil_{ x}$,
which is a {\it finite dimensional} Hilbert space for each fixed
$V$. Then we consider the C*-algebra $\A_V=B(\Hil_V)$ which, if
$\Hil_{ 0}=\mathbb{C}^2$, is isomorphic to the set of $(2|V|)\times
(2|V|)$ matrixes with complex entries.

The family of Hilbert spaces and C*-algebras constructed in this
way are related to each other in the following easy and natural
way: if $V\subset V'$ then
$\Hil_{V'}=\Hil_V\otimes\Hil_{V'\setminus V}$ and, $\forall A\in
\A_V$, $A\otimes \I_{V'\setminus V}\in \A_{V'}$.

Moreover, the map $\gamma_{ k}(a^{(1)}_{x_1}\,a^{(2)}_{x_2}\ldots
a^{(n)}_{x_n})=a^{(1)}_{x_1+k}\,a^{(2)}_{x_2+k} \ldots
a^{(n)}_{x_n+k}$ is an automorphism for each $k$ and it represents
the {\it spatial translations}.

Finally, the {\it  local energy} is given by summing up the
interactions of all the particles inside $V$,
$$
H_V=\sum_{r}\,\sum_{x_1,\ldots,x_r\in V}\,V_r(x_1,x_2,\ldots,x_r)
$$
where $V_r$ is the r-body interaction.

\vspace{2mm}

{\bf Example 2:} {\it continuous system}

The starting point, in this case, are the Fermi or the Bose
commutation rules, given in terms of {\it smeared fields}
$\Psi[f]=\int_\mathbb{R}\,\Psi(x)\,f(x)\,dx$, where we take $f(x)\in
\Sys(\mathbb{R})\subset\Lc^2(\mathbb{R})$ for technical convenience:
$$
[\Psi[f],\Psi^\dagger[g]]_{\pm}=<\overline{g},f>,$$
$$
[\Psi[f],\Psi[g]]_{\pm}=[\Psi^\dagger[f],\Psi^\dagger[g]]_{\pm}=0,
\forall f,g\in\Sys(\mathbb{R}).
$$

Let $\Phi_0$ be the {vacuum} of the theory, i.e. a vector such that
$\Psi[f]\Phi_0=0$, $\forall f\in \Sys(\mathbb{R})$. The Hilbert
space $\Hil_V$ is the norm closure of
${\Psi^\dagger[f_1]\ldots\Psi^\dagger[f_n]\Phi_0}$, where each $f_j$
is supported in $V$. Observe that, even if $V$ is a finite volume,
dim$(\Hil_V)=\infty$. This implies that, even for finite systems,
unbounded operators may appear in the game. To avoid this unpleasant
aspect one usually introduces the following C*-algebra $\A_V=\{X\in
B(\Hil_V): [X,N_V]=0\}$, where
$N_V=\int_V\,\Psi^\dagger(x)\Psi(x)\,dx$ is the {\it number
operator}. In this way, we will only consider those observables in
$\A_V$ which are automatically {\it bounded}.

One of the relevant operators is the energy, i.e. the {\it local
hamiltonian}, which for a 2-body interaction, is:
$$
H_V=\frac{\hbar^2}{2m}\int_V\,dx\,|\nabla\Psi(x)|^2+$$
$$+\frac{1}{2}\int_V dx\int_{V}dx'
\Psi^\dagger(x)\Psi^\dagger(x')V(x,x')\Psi(x')\Psi(x),
$$
but we see that, in principle, $H_V\notin \A_V$ since $H_V$ may be
unbounded!

\vspace*{3mm} {\bf The states.} Continuing with  Haag and Kastler's
construction, we recall that the  states of $\Sigma$ are positive,
normalized linear functionals on $\A$ which, when restricted to $V$,
reduces to the states over the {\it finite} system $\Sigma_V$ and,
therefore, over the finite volume algebra $\A_V$. In other words,
they corresponds to a family of {density matrices} $\rho_V$:
$\hat\rho(A)=tr_V(\rho_VA)$ for each $A\in\A_V$,
 here $tr_V$ is the trace in $\Hil_V$. These states satisfy the following
{\it consistency condition}: $tr_V(\rho_VA)=tr_{V'}(\rho_{V'}A)$
$\forall A\in\A_V$, $V\subset V'$.

They have a physical interpretation which is given by the Ruelle,
Dell'Antonio and Doplicher theorem: these  states have zero
probability to describe an infinite number of particles in a finite
region. They are usually called in the literature {\it locally
finite states}.

\vspace{3mm}

Among all the states a particular role is played by the so called
{\it pure states}: $\rho$ is {\it pure} if it is not a convex
combination of other states, i.e. if there are no $\rho_1, \rho_2$
and $\lambda\in]0,1[$ such that
$\rho=\lambda\rho_1+(1-\lambda)\rho_2$. Their relevance is due to
the fact that, as we will discuss in the following, they are related
to the pure thermodynamical phases of a certain physical system.

Lanford and Ruelle introduced the notion of {\it states with short
range correlations} (SRCS), which are given as follows: let $B$ be a
local bounded observable, $\epsilon$ a positive number. Then $\rho$
is a SRCS
 if there exists a bounded region $\Lambda$ such that, for all $A$
bounded and localized outside $\Lambda$, then
$$\left|\rho(AB)-\rho(A)\rho(B)\right|\leq \epsilon \|A\|.$$
These states are related to the pure ones. Indeed, in 1969, Ruelle
proved that {\it each pure state  is automatically a SRCS}.

A weaker requirement is the so-called {\it asymptotic abelianess}:
$\rho$ satisfies the asymptotic abelianess if, for each local $A,B$,
we have
$$
\left|\rho(A\gamma_j(B))-\rho(A)\rho(\gamma_j(B))\right|\rightarrow
0,
$$
when $|j|\rightarrow\infty$, which means that the quantity
$\rho(A\gamma_j(B))$ factorizes whenever the two observables $A$
and $\gamma_j(B)$ are localized in  regions of the space which are
far away from one another.

\vspace{3mm}

{\bf The dynamics.} The next step in our analysis is related to the
description of the {\it time evolution} of the physical system
$\Sigma$. This is obtained from the dynamics of $\Sigma_V$ in
Heisemberg representation as follows:

first we define the time evolution of the element $A$ in $\A_V$ in
the volume $V$ as follows:  $\A_V\ni A\rightarrow
\alpha_V^t(A):=e^{iH_Vt/\hbar}Ae^{-iH_Vt/\hbar}$.

secondly we use $\alpha_V^t(A)$ to define $\alpha^t(A)$ as follow
$\alpha^t(A)=\tau-\lim_V \alpha_V^t(A)$, where $\tau$ is a {\it
reasonable} topology of $\A$, i.e. a topology usually related to
$\Sigma$ itself. Possible topologies are the following:

for short range interactions and discrete systems $\tau$ is usually
the {\it uniform} topology, \cite{HHW};


for long range interactions it is known that $\alpha_V^t$ is {not
$\|.\|-$converging}: a possible alternative for $\tau$ is the {\it
strong} topology (restricted to a {\it relevant family of states}).
This different topology has been used in many papers, among which
\cite{bagmor,morstr,thi} and references therein. In this case a
state $\rho$ must be chosen in such a way that
$$\rho(\alpha_V^t(A))\rightarrow \rho(\alpha^t(A))=:\rho_t(A),$$

and this limit defines the time evolution of the state $\rho$,
$\rho_t$, by means of  $\rho_t(A):=\rho(\alpha^t(A))$. It is clear
that the existence of (sufficiently many) such $\rho$'s has to be
checked in each model.

    \vspace{2mm}

{\bf The symmetry.} It is now possible to introduce the concept of
symmetry: an automorphism of $\A$, $\gamma$, is a {\it symmetry} of
the system $\Sigma$ if $\alpha^t(\gamma(A))=\gamma(\alpha^t(A))$ and
is a {\it local symmetry} if $\gamma:\A_V\rightarrow\A_V$ and if
$\gamma(H_V)=H_V$. We have already seen an example of a local
symmetry at the beginning of this section, when speaking of the
Ising model.

Moreover, the automorphism $\gamma$  is a {\it  symmetry of the
state} $\rho$ if $\rho_\gamma(A):=\rho(\gamma(A))=\rho(A)$,
$\forall A\in\A$.

    \vspace{2mm}

{\bf Representations and GNS-construction.} A crucial notion, also
in view of its physical applications, is that of a *-{\it
representation} of a
*-algebra. This is essentially a map $\pi:\A\rightarrow B(\Hil)$,
for a certain $\Hil$, which preserves the algebraic structure of
$\A$:
$$\pi(A+B)=\pi(A)+\pi(B), \hspace{3mm} \pi(\lambda
A)=\lambda \,\pi(A),$$ $$ \pi(AB)=\pi(A)\pi(B),\hspace{5mm}
\pi(A^*)=\pi(A)^*.$$ It is clear that $\pi(\A)$ is a *-algebra as
well.

\vspace{2mm}

It is a well known fact that any state $\rho$ over an abstract
C*-algebra $\A$ produces a unique (but for equivalence) triplet
$(\Hil_\rho,\pi_\rho,\Omega_\rho)$, where $\Hil_\rho$ is an Hilbert
space, $\pi_\rho$ is a representation (in the sense discussed above)
and $\Omega_\rho$ is a cyclic vector of $\Hil_\rho$, {  i.e.
$\pi_\rho(\A)\Omega_\rho$ is dense in $\Hil_\rho$}. Moreover we
have, $\forall A\in\A$,
$$\rho(A)=<\Omega_\rho, \pi_\rho(A)\Omega_\rho>.$$ Also, $\pi_\rho$
is irreducible if and only if $\rho$ is pure.


We can give here the sketch of the proof:  $\A$ becomes a
pre-Hilbert space wrt the following positive semidefinite scalar
product: $(A,B)=\rho(A^*B)$. Let ${\cal I}_\rho=\{A\in\A:
\,\rho(A^*A)=0\}$. This is a left ideal of $\A$ ($A\in {\cal
I}_\rho, X\in\A$ then $XA\in {\cal I}_\rho$). We introduce the
{\it equivalence classes}: $[A]={A+I, \,I\in {\cal I}_\rho}$ which
produce a complex vector space when equipped with the following
operations: $[A]+[B]=[A+B]$, $[\lambda A]=\lambda [A]$. In this
way the set $\{[A],\,A\in\A\}$ is equipped with a positive
definite scalar product: $<[A],[B]>=(A,B)=\rho(A^*B)$. If we
complete $\{[A],\,A\in\A\}$ wrt the norm inherited from $<,>$ we
get our Hilbert space $\Hil_\rho$.

The representation $\pi_\rho$ is defined by $\pi_\rho(A)[B]:=[AB]$,
while the cyclic vector $\Omega_\rho$ is simply $\Omega_\rho=[\I]$.
Notice that, incidentally, $\pi_\rho$ is a {  bounded
representation} since $\|\pi_\rho(A)[B]\|\leq \|A\|\|B\|$, for each
$A,B\in \A$. This means that, $\forall A\in\A$, then $\pi_\rho(A)\in
B(\Hil_\rho)$.


{\bf Remarks:--} (1)  The first obvious remark is that GNS
representations generated by different states need not be unitarily
equivalent!

(2)   Each (GNS) representation corresponds to a {\it phase} of
the physical system. In particular,
 GNS representations generated by pure states correspond to
pure phases \cite{ruelle}.

(3) States which are only {\it locally different} are {\it
macroscopically indistinguishable}: all the macroscopic observables
have the same expectation values. For instance, if we go back to the
Ising model, it is clear that
$$\lim_{V,\infty}\,<\Phi_0^{(+1)},\pi^{(+1)}(\sigma_V^3)\Phi_0^{(+1)}>=1=
\lim_{V,\infty}\,<\pi^{(+1)}(A)\Phi_0^{(+1)},\pi^{(+1)}(\sigma_V^3)\pi^{(+1)}(A)\Phi_0^{(+1)}>,$$
for any strictly localized $A\in\A$, \cite{sewbook1}. Then they
produce unitarely equivalent GNS representations. In \cite{sewbook1}
the author says that two locally different states belong to the same
island or to a given folium, see \cite{sewbook2}.

This has a clear physical interpretation:  {equal values of the
macroscopic observables (the so-called
 {\it order
 parameters}) label unitarily equivalent  representations, which are interpreted as the
 same phase of the matter.}
In other words: two different phases of the matter correspond to
 two representations in which some macroscopic observable assumes different values.

(4) An interesting result is the following: even if the algebraic
dynamics for $\Sigma$ cannot be given in an hamiltonian form,
nevertheless, under certain assumptions on $\Sigma$, the dynamics in
each representation $\pi_\rho$ is {\it hamiltonian}: there exists a
s.a. operator $\hat H_\rho$ such that, $\forall A\in\A$,
$$
\frac{d}{dt}\,\alpha_\rho^t(\pi_\rho(A))=i[\hat H_\rho,
\alpha_\rho^t(\pi_\rho(A))],
$$
see \cite{sewbook1} and references therein. $\hat H_\rho$ is what is
often called in literature {  \it the effective hamiltonian}.

This result has a clear physical interpretation: different phases of
$\Sigma$ may have, and they usually have, different dynamical
behaviors, and this is reflected in the different possible
expressions for $H_\rho$.

\subsection{Non-zero temperature}

We devote this subsection to some brief  considerations on {\it
equilibrium states} and to the associated {\it   phase structure},
considering separately the cases of quantum systems with finite
and infinite degrees of freedom.

\vspace{2mm}

Let us start considering finite systems. In this case we can prove
that the following are equivalent:

(i) $\rho$ is a Gibbs state corresponding to the trace class
operator $\hat\rho=\frac{e^{-\beta H_V}}{tr_V(idem)}$, where
$\beta^{-1}=kT$;


(ii) $\rho$ minimizes the free energy functional $\hat
F_V(\rho)=tr_V(\rho H_V+\beta^{-1}\rho\,\log(\rho))$;


(iii) $\rho$ is a {KMS} (i.e. {\it Kubo-Martin-Schwinger}) state at
the corresponding inverse temperature $\beta$, i.e., roughly
speaking, if $A,B\in\A$, $\rho(A_tB)=\rho(BA_{t+i\hbar \beta}),$
$\forall t$

This equivalence has an immediate consequence: for each temperature
there exists an unique equilibrium state, and, therefore,  an unique
associated GNS representation. This means that, for such a finite
system, there exists a single thermodynamical phase of $\Sigma$ at
each given temperature.

\vspace*{2mm} The situation is completely different for infinite
systems. For these systems the role of the thermodynamical limit is
crucial, and not only for the existence of the {\it time evolution},
as we will discuss in a moment.

In order to keep the analysis simple, it is convenient to make the
following assumptions on the finite volume hamiltonian $H_V$:
\begin{enumerate}
\item first we require that $H_V$ is such that
$H_{V_1\cup V_2}-H_{V_1}-H_{V_2}$ is a surface effect. This
condition holds, for instance, for short range forces;
\item there exists $c>0$ such that $\|H_V\|\leq c |V|$.
\end{enumerate}
These assumptions imply, first of all, that
$\alpha^t(A)=\|\,\|-\lim_{V\nearrow}\,\alpha_V^t(A)$.

Secondly, let us define the following functionals $$ \left\{
    \begin{array}{ll}
    E_V(\rho_V)=tr_V(\rho_VH_V),\\ S_V(\rho_V)=-k
\,tr_V(\rho_V\,\log(\rho_V)),\\
F_V(\rho_V)=E_V(\rho_V)-TS_V(\rho_V)\\
\end{array}
  \right.
$$
These are called the local {\it energy}, the {\it entropy} and the
{\it free energy} functionals. In particular, the entropy functional
satisfies the following crucial inequalities, useful to compute the
thermodynamical limits of some functional related to $S_V(\rho_V)$:
for each $V_1\cap V_2=\emptyset$ then $$ S_{V_1\cup V_2}(\rho)\leq
S_{V_1}+ S_{V_2}.$$ This is the so-called {\it subadditivity}
property of the entropy. The {\it strong subadditivity} property,
proved later by Lieb and Ruskai, also holds: for each $V_1, V_2$,
the following inequality is satisfied:
$$ S_{V_1\cup V_2}(\rho)+S_{V_1\cap V_2}(\rho)\leq S_{V_1}+
S_{V_2}.$$

{\bf Remark:--} In view of the pedagogical nature of this paper, it
may be of some interest to comment briefly about the definition of
the entropy functional. As a matter of fact this can be seen as a
quantum counterpart of a concept arising from information theory,
where one consider the entropy as a {\it mean surprise}. Let us show
how:

let us consider a set of $M$ {\it  elementary events}
$\{\E_1,\ldots,\E_M\}$ and let $p_j$ be the probability that the
event $\E_j$ occurs: of course we have $p_j\geq 0$ and
$\sum_{j=1}^M\,p_j=1$. Let further $u_j=-\log(p_j)$ be the {\it
surprise} related to $\E_j$: with this definition it appears clear
that, if we are sure that $\E_j$ is going to occur ($p_j\simeq
1^-$), then there is no surprise at all (and indeed we have
$u_j\simeq 0$). On the other way, if $\E_j$ is extremely rare
($p_j\simeq 0^+$), then the surprise is very large (and we find
$u_j\simeq \infty$).

The {\it  mean surprise} is defined as
$$MS=\frac{\sum_{i=1}^M\,N_i\,u_i}{\sum_{i=1}^M\,N_i}=\sum_{i=1}^M\,p_i\,u_i=-\sum_{i=1}^M\,p_i\,\log(p_i),$$
where $p_i=N_i/N$, which is exactly the Shannon expression of the
entropy. The generalization from the classical to the quantum
entropy gives rise to the definition above.

\vspace{3mm}

The assumptions for $H_V$ and the  subadditivity of the entropy,
imply that the following {\it global density functionals}
$$ e(\rho)=\lim_{V\nearrow}\,\frac{E_V(\rho_V)}{|V|}, \,
s(\rho)=\lim_{V\nearrow}\,\frac{S_V(\rho_V)}{|V|}, \,
f(\rho)=\lim_{V\nearrow}\,\frac{F_V(\rho_V)}{|V|}
$$
exist, as well as the following {\it incremental functionals}
 $$ \left\{
    \begin{array}{ll}
    \Delta E(\rho|\rho')=\lim_{V\nearrow}\,\left(E_V(\rho'_V)-E_V(\rho_V)\right),\\
    \Delta S(\rho|\rho')=\lim_{V\nearrow}\,\left(S_V(\rho'_V)-S_V(\rho_V)\right),\\
\Delta F(\rho|\rho')=\lim_{V\nearrow}\,\left(F_V(\rho'_V)-F_V(\rho_V)\right),\\
\end{array}
  \right.
$$
where $\rho'$ is a {\it local} modification of $\rho$, i.e. a
state which differs from $\rho$ only on a volume of finite size.

The role of these functionals is crucial in the analysis of the
equilibrium states.  A state $\tilde\rho$ is {\it globally
thermodynamically stable} (GTS) if it is invariant under
translations and if it minimizes $f(\rho)$. It is {\it locally
thermodynamically stable} (LTS) if $\Delta F(\tilde\rho|\rho')\geq
0$ for all $\rho'$, local modification of $\tilde\rho$. Then,
\cite{sewbook1}, it is proved that a GTS state is a LTS state, while
an LTS state which is invariant under translations is also a GTS
state for systems with short range interactions.

Again, this result has a physical interpretation, which can be
deduced also from the explicit solution of some easy physical
models: a GTS state is an equilibrium state. The LTS states are, for
systems with long range interactions, only {\it metastable} states
(i.e. states with a long mean life and {\it good} thermodynamical
properties). Obviously, they are also true equilibrium states under
the above assumptions.

Another interesting result relates the LTS and the KMS states: they
are exactly the same objects, \cite{sewbook1}! This implies, of
course, that a KMS state for an infinite system is not an
equilibrium state, in general, but only a metastable state. This is
different from what happens for finite systems, as we have already
pointed out.

It may be worth, at this stage, recalling some general results on
the KMS states. Let $\Sigma$ be an finite system. Then, as we have
already said, a state $\rho$ over a C*-algebra $\A$ is a KMS-state
at an {\it inverse temperature} $\beta$ (briefly, a $\beta$-KMS
state in the following) if, for all observables $A,B$ and for all
$t\in\mathbb{R}$,
$$\rho(A_t\,B)=\rho(B\,A_{t+i\hbar\beta})$$
For infinite systems this definition does not work, in general,
since, e.g., $A_{t+i\hbar \beta}$ may make no sense. Also, and even
more substantial, the time evolution may not exist even for real
time. Therefore, in reference \cite{HHW},  a different definition
was proposed:

\vspace{2mm}

$\rho$ is a $\beta$-KMS state if for each $A,B\in\A$ there exists
a complex function $F_{AB}(z)$ which is analytical in the strip
$\Im(z)\in[0,\hbar \beta]$, continuous on the boundaries, and is
such that
$$
F_{AB}(t)=\rho(BA_t),\hspace{1cm} F_{AB}(t+i\hbar\beta)=\rho(A_tB)
$$

The physical interpretation of  KMS-states are well established by
some explicitly solvable quantum models: a $\beta$-KMS state,
$\rho_\beta$, is nothing but a {\it reservoir} at a temperature
$T=\frac{1}{k\beta}$. Indeed, given $\Sigma$ described by
$\rho_\beta$ and weakly coupled with a finite system $\Sys$, in the
limit $t\rightarrow\infty$, one can prove that, independently of the
details of the $\Sigma-\Sys$ interaction and of the initial state of
$\Sys$, this is described by a Gibbs state corresponding to the same
inverse temperature $\beta$ of $\Sigma$.

\vspace*{2mm}

What makes the difference between finite and infinite systems is now
the following remark: while for a given temperature the equilibrium
state of a finite system is uniquely fixed by any of the three
equivalent requirements discussed above,  an infinite system
$\Sigma$ may possess more than one GTS state at the same
temperature. Examples of different GTS states may be constructed as
limits of a Gibbs state (for those thermodynamical conditions)
corresponding to different boundary conditions. This result has a
mathematical interpretation which is quite simple: while the
potential $\hat F_V(\rho)$ is convex, and therefore admits an unique
minimum,  the free energy density functional  $f(\rho)$ is affine,
so that more than a single minimum may be achieved.

If this is the case, $\Sigma$ admits different thermodynamical
phases under the same thermodynamical conditions, each corresponding
to a different GTS state. We say that the system possesses {\it
macroscopic degeneracy}: these different equilibrium states  (and
the related physical phases) are labeled by the (different) values
of some {\it macroscopic observables} (like the magnetization in the
case of ferromagnetic materials), \cite{sewbook1}.

\vspace{3mm}

This fact has several related consequences, which we here list and
comment briefly, referring to specific textbooks for a deeper
analysis:

The first consequence of  the algebraic approach discussed so far,
and the possibility of having a macroscopic degeneracy for
infinitely extended systems, is that it provides a nice framework to
analyze coexisting phases of such systems and, as a consequence, to
 discuss easily the occurrence of phase transitions.

A second consequence, which is deeply connected with the first
one, is that we can use this approach to discuss the so called
{\it spontaneous breaking of a symmetry}:

suppose that $\Sigma$ has a local symmetry $\gamma$ and let
$\Delta=\{\rho\in\A': \rho \mbox{ is GTS}\}$, be the set of GTS
states. Then, since necessarily we have $f(\rho)=f(\rho_\gamma)$, it
follows that for any $\rho\in\Delta$ also $\rho_\gamma\in\Delta$:
the symmetry $\gamma$ maps $\Delta$ into itself.

From that we see that, if $\Delta=\{\rho_1\}$ consists of a single
element, $\gamma$ is necessarily a symmetry of $\rho_1$: {\it   the
symmetry is unbroken}. In other words, in this case it is clear that
$\rho_1=(\rho_1)_\gamma$.

If, on the contrary, $\Delta=\{\rho_1,\ldots,\rho_n\}$, then, in
general, we can only say that $(\rho_i)_\gamma=\rho_j$, for some
$i$ and $j$ not necessarily equal: if this is the case, then {\it
the symmetry is spontaneously broken}.

\vspace{2mm}

 {\bf Example (T=0 Ising model)}: for this model $H_V$ is
invariant under spin reversal ($\sigma^3_j\rightarrow -\sigma^3_j$),
which is therefore a local symmetry, since it also maps $\A_V$ into
itself, but the two (transactionally invariant) ground states
associated to the magnetization $m=\pm 1$ are, clearly, no longer
invariant: they are mapped into each other by the symmetry. To be
more explicit, the vectors $\psi^{(\pm 1)}$ are the cyclic vectors
related, via a standard GNS construction, to the two states
$\rho_{\pm}$ over the quasi-local C*-algebra of the discrete system,
which are characterized by the values of the following limits:
$$
\lim_{N\rightarrow\infty}\rho_{\pm}\left(\frac{1}{2N+1}\sum_{j=-N}^N\,\sigma^3_j\right)=\pm
1.
$$
In this example it is clear that $\Delta=\{\rho_+,\rho_-\}$.

\vspace*{3mm}

Whenever a system exhibits a spontaneously symmetry breaking, a
related result on the spectrum of the theory can be deduced by
making use of the {\it non relativistic Goldstone's theorem}.

Roughly speaking, this theorem says the following: suppose that
the symmetry $\gamma_\lambda$ is generated by a
 local charge $$Q_R(t)=\int_{|\vec
x|\leq R}\,j_o(\vec x,t)\,d^3x,\vspace{.4cm}\mbox{and
}\vspace{.4cm}\gamma_\lambda(A)= \|.\|\lim_{R,\infty}
e^{iQ_R\lambda}Ae^{-iQ_R\lambda},$$ and suppose that
$\gamma_\lambda$ commutes with the time translations $\alpha^t$.
Then, if $\gamma_\lambda$ is spontaneously broken, i.e. if for some
$A\in\A$, $\lim_{R,\infty}<[Q_R,A]>_{\psi_0}\neq 0$, then the energy
spectrum cannot have a gap above the ground state.

This theorem has proved to be a very important tool both in
condensed matter and in quantum field theory.

\vspace*{2mm}

We want to end this list of results related to our algebraic
approach by mentioning a very interesting relation between KMS
states and the Tomita-Takesaki theory, which enriches the list of
relevant results which can be easily proven using the C*-algebraic
picture of $QM_\infty$.

We start recalling very briefly few facts on this theory:

first let us recall that a {\it   von Neumann algebra} (VNA) is a
 selfadjoint (i.e. closed with respect to the adjoint) subset of $B(\Hil)$, ${\cal
M}\subset B(\Hil)$, such that ${\cal M}={\cal M}''$. Here ${\cal
M}'=\{X\in B(\Hil),\,\, [X,A]=0\,\,\forall A\in{\cal M}\}$ is called
the {\it commutant} of ${\cal M}$ and ${\cal M}''$, which is
constructed in the same way, is its {\it bicommutant}. Equivalently,
${\cal M}\subset B(\Hil)$ is a VNA if it is weakly closed or
strongly closed.

 This implies that every VNA is a C*-algebra (indeed, if ${\cal M}$
 is
weakly closed then ${\cal M}$ is automatically uniformly closed),
while not any C*-algebra is a VNA (e.g. $C_0(X)$).

Tomita-Takesaki's theorem is given for $\sigma-$finite VNAs, for
which a cyclic and separating vector surely exists. We recall that
a vector $\varphi\in\Hil$ is said separating for $\cal M$ if,
 $X\varphi=0$ for $X\in{\cal M}$ is equivalent to $X=0$. Let $\Omega$ be
 a cyclic and separating
vector, and let $S_0$ and $F_0$ be the densely defined operators
$$S_0 A\Omega=A^*\Omega, \,\,\,\,F_0 A'\Omega=A'^*\Omega,\,\, \forall
A\in{\cal M},\,\forall A'\in{\cal M}'.$$ These operators are
closable, $S=\overline{S_0}$, $F=\overline{F_0}$, and the polar
decomposition of $S$, $S=J\Delta^{1/2}$, produces the {\it modular
conjugation $J$} and the {\it   modular operator $\Delta$} {\it
associated to} $({\cal M}, \Omega)$.

Many results have been discussed in the literature concerning $J$
and $\Delta$, but since they have no role here, we will not give
further details, referring to \cite{br} for more information. We
just want to mention here the following result:

{\bf Tomita-Takesaki theorem:} With the above definitions we have
$$J{\cal M}J={\cal M}',\mbox{ and }\hspace{1cm} \Delta^{it}{\cal M}\Delta^{-it}={\cal M},$$
for all $t\in\mathbb{R}$.

\vspace{2mm}

It is not very hard to show now that all the KMS states can be
used to generate a modular structure in the sense of
Tomita-Takesaki:

let $\rho$ be a $-1$-KMS state (i.e. a KMS state corresponding to
$\beta=-1$. This is what it is usually called simply a KMS state).
This state generates a GNS representation $(\Hil_\rho, \pi_\rho,
\Omega_\rho)$. Let $U_\rho(t)$ the unitary operator which implements
$\alpha^t$ in this representation. Then $\Omega_\rho$ is cyclic and,
{  since $\rho$ is KMS}, is also separating for $\pi_\rho(\A)''$,
i.e. $\pi_\rho(X)\Omega_\rho=0$ implies that $\pi_\rho(X)=0$,
\cite{br}.

Then we are in the assumptions of  Tomita-Takesaki's construction,
so that we can introduce a {\it   modular conjugation $J_\rho$} and
a {\it modular operator $\Delta_\rho$} associated to
$(\pi_\rho(\A)'',\Omega_\rho)$. Calling $H_\rho$ the generator of
$U_\rho(t)$, we find that $$\Delta_\rho=e^{ H_\rho}.$$ In other
words: given a system $\Sigma$ an {\it effective hamiltonian} surely
exists in any representation GNS-constructed by a given KMS-state.


\vspace{3mm} It may be worth stressing that these are only few
results which can be obtained within an algebraic frameworks. More
results on phase transitions, applications to quantum field theory,
statistical mechanics etc. can be found in many specialized
textbooks,  among which we only cite
\cite{thirringbook,br,br2,sewbook1,sewbook2}.

\subsection{A list of problems}

Instead of giving more results on this {\it canonical scheme}, we
devote the last part of this section to discuss some limits which
are, in our idea, intrinsic with the approach discussed so far, and
which suggest the construction of a slightly generalized algebraic
framework. These conclusions  are based on a simple remark: the main
results which have been given in this section are obtained under
some requirements, which may  not be necessarily satisfied in many
relevant conditions. For instance, we have assumed that
 {\it the norm of the local hamiltonian
$H_V$ does not grow faster than $|V|$: $\|H_V\|\leq c |V|$}.

However, this is not always true: actually, it is false quite often!
For instance, this inequality is violated already by a gas of free
bosons, for which $H_V=\sum_{j\in V}a_j^\dagger a_j$, since each
creation and annihilation operator  is such that
$\|a_j\|=\|a_j^\dagger\|=\infty$.

This same condition is satisfied, on the contrary, by a gas of free
fermions, for which, however, $dim(\Hil_V)<\infty$. This is one of
the reasons why, in the analysis of an open system, the free gas of
bosons which constitutes the reservoir is frequently replaced by a
gas of fermions. We will come back on this point in a moment.

The second assumption considered above is that {\it the interactions
are  short ranged}.

However this is not the case in many situations. For instance, the {
Coulomb interaction} is long ranged, while in the { mean field }
models the {\it real forces} are replaced by interactions with an
infinite range: this means that, given two particles localized in
$i$ and $j$, they feel the same strength independently of the
difference $|i-j|$. However, many results can
be obtained even under these conditions. In particular we find that:\\
{$\bullet$} {   $\alpha_V^t$ is not norm convergent to an algebraic
dynamics $\alpha^t$}, but, as we have already sketched before, we
can (often) find a different topology which makes of $\alpha_V^t(A)$
a Cauchy sequence for each (or for many) $A\in\A$, see
\cite{bagmor,thi,DS} just to cite few authors.\\
{$\bullet$} {We have already mentioned that, in general,   KMS
states are not equilibrium states (i.e. GTS states). Moreover, they
are not even limits of Gibbs state}; indeed, in \cite{haag}, it is
stated the following result, which we repeat here in a simplified
version:

if $\alpha_V^t$ is uniformly convergent to $\alpha^t$ and if,
calling $\omega_\beta^{(V)}(A)=tr_V(\rho_{\beta,V}A)$, with
$\rho_{\beta,V}=\frac{e^{-\beta H_V}}{tr(idem)}$, the limit
$\lim_V\omega_\beta^{(V)}(A)=\omega_\beta(A)$ exists for each
$A\in\Ao$, then:

\vspace{1mm}

a. $\omega_\beta$ is a $\beta$-KMS state wrt $\alpha^t$;

b. $\omega_\beta$ is associated to a modular operator and a
modular conjugation (in the sense of Tomita-Takesaki).

{\bf Remark:--} it is clear that we are requiring here the {\bf
uniform convergence} of $\alpha_V^t$, which, as we have just seen,
is violated for long range interactions, and the {\bf existence of
the limit} of
$\omega_\beta^{(V)}(A)$, which is not ensured a priori!! Therefore we cannot
conclude that limit of Gibbs states are surely KMS states.\\
{$\bullet$} {   surface effects become volume effects, so that {\it
variables at infinity} (i.e. completely delocalized operators)
 appear in the dynamics of strictly  localized operators. These are related
 to the {\it order parameters} used to describe different phases, \cite{morstr} and references therein;}\\
{$\bullet$}   the density functionals $e(\rho)$ and $f(\rho)$ do not
necessarily exist,
since in the proof of their existence, the assumption that the forces are short ranged is crucial, see \cite{sewbook1};\\
{$\bullet$} {   the Goldstone's theorem holds only in a modified
form} \cite{morstr}.

We see that a new world appears whenever the interactions appearing
in the physical system modify their range. We also observe that many
things can be said but many other aspects are still to be clarified.

\vspace*{1mm}

A third assumption which is usually somehow hidden in the
C*-algebraic approach to $QM_\infty$ is related again to the
presence on the (almost) unavoidable unbounded operators. Consider,
for instance, the position and momentum operators $\hat q$ and $\hat
p$. As we have already mentioned, they satisfy the following
commutation relations: $[\hat q,\hat p]=i\,\id$ (in convenient
units) and, as a consequence, it is an easy exercise to check that
at least one of them must be unbounded. Actually, it is well known
that they are both unbounded. In the literature three possible ways
to deal with unbounded operators have been proposed: the first one
consists in restricting the action of the operators on some
(possibly dense) subset of a given Hilbert space, a sort of {\it
common domain} of all the operators. A second possibility is to
exponentiate these unbounded and self-adjoint operators in order to
define unitary (and therefore bounded) operators. The original
operators can be recovered by taking suitable derivatives of the
unitary maps on certain relevant sets of vectors. A third
possibility is the following: we could replace, say, the operator
$\hat p$ with a bounded operator $\hat p_N$ whose spectrum coincides
with the one of $\hat p$ inside a compact interval $[-N,N]$, and is
zero outside this set. It is clear that $\hat p_N$ is bounded and,
as $N\rightarrow\infty$, approaches $\hat p$ in some sense. Then one
considers only those states $\omega_N$ on $\A$, $N$-depending as
well, such that $\omega_N(\hat p_N)$ converges in the limit
$N\rightarrow\infty$ to some specific quantity, which in some sense
represent the mean value of the original operator $\hat p$ on a
state which can be interpreted as the {\it limit} of the family of
states $\omega_N$. An example of this procedure can be found in
\cite{as}.

\vspace{2mm} However, quite often this is not enough. As an example,
we cite the {  Lindblad expression} for the generator $L$ of a
completely positive semigroup (describing the time evolution of a
quantum open system). These structures play a very important role in
the analysis of order-disorder transitions out of equilibrium,
\cite{sewbook1,sewbook2}.

More in details, let $\A$ and $\B$ be C*-algebras. We recall that
a map $f:\A\rightarrow\B$ is {\it positive} if $f(A)>0$ for each
$A>0$. It is {\it completely positive}, CP, if, for any finite
matrix algebra ${\cal M}$, the mapping $f\otimes
\id:\A\otimes{\cal M}\rightarrow\B\otimes{\cal M}$ is positive.

Examples of completely positive maps are the following: (1) the
automorphisms of C*-algebras are CP; (2) let ${\cal K}\subset\Hil$
be both Hilbert spaces and $P$ the projection operator from $\Hil$
into $\cal K$. Then $f(A)=PAP$ is CP.

A {\it  quantum dynamical semi-group} is a set $\{T_t: \,t\geq
0\}$ of completely positive, identity preserving  maps of $\A$
such that $T_sT_t=T_{s+t}$ for all $s,t\geq 0$ and $T_0=\1$. If
$T_t$ is normwise continuous in $t$ for all $A\in\A$, then there
exists an infinitesimal generator $L$ defined by the formula
$$
\frac{d}{dt}T_tA=LT_tA=T_tLA, \hspace{5mm} \forall A\in\A
$$
Lindblad proved that, if $\A=B(\Hil)$ (for some $\Hil$),  $L$ has
necessarily the following expression:
$$ LA=i[H,A]+\sum_j\left(V_j^*AV_j-\frac{1}{2}\{V_j^*V_j,A\}\right),
$$
where $H$ is self-adjoint and $V_j, \sum_j\,V_j^*V_j\in\A$.

Very few results exist  for unbounded operators,
\cite{fagn,cheb,sewbag}, mainly because of the following technical
difficulty: if $T_t$ is a semigroup (and not a group) it follows
that $T_t(AB)\neq T_t(A)T_t(B)$. For this reason even if we can
introduce a cutoff in the system (so that all the operators we get
are bounded), no known general result on perturbation of generators
can be used because it will contrast with the final operation of
removing the cutoff!

\vspace{1mm}

As we have already mentioned, to avoid this kind of difficulties, in
many models the boson reservoir is replaced by a fermionic one, as
for instance, in \cite{MB} for the open BCS-model, changing a
realistic into a, somehow, toy model. This suggests that an
alternative procedure should be considered, and this will be the
contain of the next sections.

\section{Algebras of unbounded operators}

In this section we will briefly introduce different examples of
what we generically call {\it algebras of unbounded operators},
giving only those mathematical results and definitions which are
relevant for our purposes. A much deeper analysis of these aspects
can be found, for instance, in \cite{aitbook} or in \cite{schu}.

 A possible algebraic framework, which is also the main one we will work with here and in
 the next section, is the following:

  let $\A$ be a linear space,
$\Ao\subset\A$ a $^\ast$-algebra with unit $\1$ (otherwise  we can
always add it): $\A$ is a {\it   quasi $^\ast$-algebra over $\Ao$}
if

\vspace{3mm}

{\bf [i]} the right and left multiplications of an element of $\A$
and an element of $\Ao$ are always defined and linear;


{\bf [ii]} $x_1 (x_2 a)= (x_1x_2 )a, (ax_1)x_2= a(x_1 x_2)$ and
$x_1(a
 x_2)= (x_1 a) x_2$, for each $x_1, x_2 \in \A_0$ and $a \in \A$;


{\bf [iii]} an involution * (which extends the involution of $\Ao$)
is defined in $\A$ with the property $(ab) ^\ast =b ^\ast a ^\ast$
whenever the multiplication is defined.



\vspace{3mm}

A quasi  $^\ast$ -algebra $(\A,\Ao)$ is {\it   locally convex} (or
{\it   topological}) if in $\A$ a { locally convex topology} $\tau$
is defined such that (a) the involution is continuous and the
multiplications are separately continuous; and (b) $\Ao$ is dense in
$\A[\tau]$.

Let $\{p_\alpha\}$ be a directed set of seminorms which defines
$\tau$. The existence of such a directed set can always be assumed.
We can further  also assume that $\A[\tau]$ is {\it complete}.
Indeed, if this is not so, then the $\tau$-completion
$\tilde\A[\tau]$ is again a topological quasi *-algebra over the
same
*-algebra $\Ao$.

\vspace{2mm}

One may ask why  these structures are related to unbounded
operators. This can be understood in a simple way just remarking
that, if $a$ and $b$ are unbounded operators, then $ab$ and $ba$ do
not exist in general. But if $x$ is a third bounded operator, then
$xa, ax, bx$ and $xb$ are all well defined, at least if the range of
$x$ is contained in the domain of $a$ and $b$. This is reflected by
the fact that $\A$ is not a
*-algebra, while $\Ao$ is, but only a {\it quasi
*-algebra}: not all its elements can be mutually multiplied, but we
can safely multiply an element of $\A$ with an element of $\Ao$.

This abstract argument can be made more explicit by the next
example, which shows explicitly that some concrete realization of
{ $(\A,\Ao)$ contains unbounded operators}.


{\bf    Example:} Let $\Hil$ be a separable Hilbert space and $N$ an
unbounded, densely defined, self-adjoint operator. Let $D(N^k)$ be
the domain of the operator $N^k$, $k\in \N$, and $\D$ the domain of
all the powers of $N$:  $ \D \equiv D^\infty(N) = \cap_{k\geq 0}
D(N^k). $ This set is dense in $\Hil$. Let us now introduce
$\Lc^\dagger(\D)$, the *-algebra of all the {  closable operators}
defined on $\D$ which, together with their adjoints, map $\D$ into
itself. Here the adjoint of $X\in\Lc^\dagger(\D)$ is {
$X^\dagger=X^*_{\restr \D}$}. \footnote{We need to introduce a map
which, given an element $X\in\LL^\dagger(\D)$, produces another
element $X^\dagger\in\LL^\dagger(\D)$. The most natural choice,
which is clearly $X^\dagger\equiv X^*$, is  only compatible with
$\LL^\dagger(\D)=B(\Hil)$, i.e. with $N$ bounded, which is not what
we want. Recalling that $D(X^*)\supseteq\D$, it is clear that
$X^*_{\restr \D}$ is well defined. Further one can prove that
$\dagger$ has the properties of an involution and maps
$\LL^\dagger(\D)$ into itself.}

\vspace{3mm}

In $\D$ the topology is defined by the following $N$-depending
seminorms: $\phi \in \D \rightarrow \|\phi\|_n\equiv \|N^n\phi\|,$
 $n\in \mathbb{N}_0$, while the topology $\tau_0$ in $\Lc^\dagger(\D)$ is introduced by the seminorms
{\normalsize$$ X\in \Lc^\dagger(\D) \rightarrow \|X\|^{f,k} \equiv
\max\left\{\|f(N)XN^k\|,\|N^kXf(N)\|\right\},\vspace{-2mm}$$} where
$k\in\mathbb{N}_0$ and   $f\in\C$, the set of all the positive,
bounded and continuous functions  on $\mathbb{R}_+$, which are
decreasing faster than any inverse power of $x$:
$\Lc^\dagger(\D)[\tau_0]$ is a {   complete *-algebra}.

It is clear that $\Lc^\dagger(\D)$ contains unbounded operators.
Indeed, just to consider the easiest example, it contains all the
positive powers of $N$. Moreover, if $N$ is the closure of
$N_o=a^\dagger\,a$, with $[a,a^\dagger]=\id$, $\Lc^\dagger(\D)$ also
contains all positive powers of $a$ and $a^\dagger$.
    \vspace{2mm}

Let further {  $\Lc(\D,\D')$} be the set of all continuous maps from
$\D$ into $\D'$, with their topologies (in $\D'$ this is the {
strong dual topology}, see Appendix 2), and let $\tau$ denotes the
topology defined by the seminorms $$ X\in \Lc(\D,\D') \rightarrow
\|X\|^{f} = \|f(N)Xf(N)\|,$$ $f\in\C$. Then $\Lc(\D,\D')[\tau]$ is a
{ complete vector space}.

In this case { $\Lc^\dagger(\D)\subset\Lc(\D,\D')$} and the pair
$$(\Lc(\D_,\D')[\tau],\Lc^\dagger(\D)[\tau_0])$$ is a {\it concrete realization} of a locally convex quasi
*-algebra.

\vspace{2mm}

{\bf Remark:} let us now suppose that $\D\equiv \Sys(\mathbb{R})$,
the set of the test functions, and $\D'=\Sys'(\mathbb{R})$. Since
$\Sys(\mathbb{R})\subset \Sys'(\mathbb{R})$, it is easy to check
that $\LL^\dagger(\Sys)\subset \Lc(\Sys,\Sys')$. Let
$\Psi(x)\in\Sys'(\mathbb{R})$. We define the map $Z_\Psi$ as
follows: $(Z_\Psi f)(x)=\Psi(x)f(x)$, $\forall
f(x)\in\Sys(\mathbb{R})$. Since $\Psi(x)f(x)\in\Sys'(\mathbb{R})$,
and since $Z_\Psi$ is continuous, we conclude that $Z_\Psi\in
\Lc(\Sys,\Sys')$. It is clear that, for instance, $Z_\Psi^2$,  does
not exist for generic $\Psi$, and this reflects the fact that
$\Lc(\Sys,\Sys')$ is not an algebra.

\vspace{3mm}

Using a quasi *-algebra is not the only possibility to include
unbounded operators in a reasonable algebraic framework. For
completeness  we briefly mention now two other possibilities which,
however, will play no role in the physical applications considered
in the next section.

We begin recalling that a {\it   partial *-algebra} \cite{antkar} is
 a complex vector space $\A$ with involution
* (with the usual properties) and a subset $\Gamma\subset (\A,\A)$
such that

$(x,y)\in \Gamma$ iff $(y^*,x^*)\in\Gamma$;

if $(x,y), (x,z)\in \Gamma$ then $(x,\lambda y+\mu z)\in \Gamma$ for
all $\lambda,\mu\in\mathbb{C}$;

if $(x,y)\in \Gamma$ then there exists an element $x\cdot y\in\A$.
This multiplication satisfies the following properties:
$x\cdot(y+\lambda z)=x\cdot y+\lambda x\cdot z$ and $(x\cdot
y)^*=y^*\cdot x^*$, $\forall (x,y), (x,z)\in \Gamma$.

\vspace{3mm}

Such a structure is a generalization of a quasi *-algebra, meaning
with that that each quasi *-algebra is also a partial *-algebra.
However, from our point of view, this appear to be  too  general to
be used in concrete applications, and for this reason is not
particularly relevant in our scheme.

\vspace{3mm}

Other examples of algebras of unbounded operators are the so-called
{\it  CQ*-algebras} \cite{bagtra1996}, which can be seen as
particular cases of topological quasi *-algebras. We do not give
here the general definition but only its simplest version with some
examples, referring to
\cite{bagtra1996,bagtralp,btt,bagtra2000,baginotra2000} for more
details.

\vspace{3mm}

A {  (proper)} CQ*-algebra is a quasi *-algebra such that:
$\Ao[\|.\|_0]$ is a C*-algebra; $\A[\|.\|]$ is a Banach space in
which $\Ao$ is dense; the two norms are related as follows: $\|x\|_0
= \max\left\{ \sup_{\norm{a}\leq 1}\norm{ax}, \sup_{\norm{a}\leq
1}\norm{xa}\right\}$, $\forall x \in \Ao.$

This is a {  natural generalization of C*-algebras:} indeed the
completion of any C*-algebra $(\Ao, \norm{\,}_0)$ with respect to a
weaker norm $\norm{\,}$ satisfying: (i) $\norm{A^\ast}=\norm{A}$,
$\forall A \in \Ao$ and (ii) $\norm{AB}\leq\norm{A}\norm{B}_0$,
$\forall A,B\in \Ao$, \noindent is a CQ*-algebra.

Let us now consider few examples of CQ*-algebras:

\vspace{3mm}

{\bf   Examples ({\it commutative cases}):}

(1) We begin with $(L^p(X,\mu),\, C_0(X))$,  where $X$ is a
compact space and $C_0(X)$ is the set of the continuous functions
on $X$ \cite{bagtralp};

(2) The second abelian example is $(L^p(X,\mu),L^\infty(X,\mu))$,
where $(X,\mu)$ is a measure space with $\mu$ a Borel measure on
the locally compact Hausdorff space $X$ \cite{bagtralp}.

\vspace{3mm}

{\bf   Examples ({\it non commutative cases}):}

(3) We first mention the  non commutative $L^p$ spaces,
\cite{btt}.

(4) A second example can be constructed as follows: let ${{\mc
H}}$ be a Hilbert space with scalar product $(.,.)$ and $S$ an
unbounded selfadjoint operator, with $S\geq \id$, with dense
domain $D(S)$. The subspace $D(S)$ becomes a Hilbert space,
denoted by ${\mc H}_{+1}$, with the scalar product $ (f,g)_{+1}
=(Sf,Sg).$ and let ${\mc H}_{-1}$ denote the conjugate dual of
${\mc H}_{+1}$. Then ${\mc H}_{-1}$ itself is a Hilbert space.
Given further
$$\A=\left\{X\in{\mc B}({\mc H}_{+1},{\mc H}_{-1}): X  \mbox{ is
compact from } \Hil_{+1} \mbox{ into } \Hil_{-1} \right\},$$
$${\mc A}_\flat=\left\{X\in{\mc B}({\mc H}_{+1}): X \mbox{ is
compact in } \Hil_{+1} \right\},$$ then $({\A}[\|.\|], *,
{\A}_\flat[\|.\|_\flat], \flat)$ is a (non proper) CQ*-algebra of
operators, whose definition can be found in \cite{bagtra2000}.

\vspace{2mm}

{\bf  Remark:--} We also want to stress that these structures have
been used in relation with Tomita-Takesaki's theory,
\cite{baginotra2000}.

\vspace{2mm}




It is not surprising that, analogously to what happens for
C*-algebras, a crucial role also from the point of view of
physical application is played by the {\it *-representations of a
quasi *-algebra}.

Let now $(\A,\Ao)$ be a quasi *-algebra, $\D_\pi$ a dense domain in
a certain Hilbert space $\Hil_\pi$, and $\pi$ a linear map from $\A$
into $\LL^\dagger(\D_\pi, \Hil_\pi)$, where
\!\!{\normalsize$$\LL^\dagger(\D_\pi, \Hil_\pi)=\{ X \mbox{ closable
in } \Hil_\pi:
 D(X)  = \D_\pi \mbox{ and } D(X^*)  \supseteq  \D_\pi\}.$$}
 This is a {\it   partial
 *-algebra}
 with the usual operations
 $X+Y$, $\lambda X$, the involution $X^\dagger=X^*_{\restr\D_\pi}$ and the weak product
 {  $X${\normalsize$\Box$}$Y\equiv X^{\dagger
 *}Y$}
 (defined whenever $Y\D_\pi\subset D(X^{\dagger *})$ and $X^\dagger\D_\pi
 \subset D(Y^{*})$. Notice that these conditions produce the definition of the set $\Gamma\subset(
 \Lc^\dagger(\D_\pi,\Hil_\pi),\Lc^\dagger(\D_\pi,\Hil_\pi))$).
Let furthermore {$$\Lc^\dagger(\D_\pi)=\{A\in
\Lc^\dagger(\D_\pi,\Hil_\pi): \, A,\, A^\dagger:
\D_\pi\rightarrow\D_\pi\}.$$} $\Lc^\dagger(\D_\pi)$ is a
*-algebra and the weak multiplication $\Box$ reduces to the ordinary
multiplication of operators.

\vspace*{3mm}

In our context a {\it   *-representation of $\A$} is a linear map
from $\A$ into { $\LL^\dagger(\D_\pi, \Hil_\pi)$} such that:

(i) $\pi(a^*)=\pi(a)^\dagger, \quad \forall a\in \A$;

(ii) if $a\in \A$, $x\in \Ao$, then $\pi(a)${\normalsize$\Box$}\!\!
$\pi(x)$ is well defined and $\pi(ax)=\pi(a)${\normalsize$\Box$}\!\!
$\pi(x)$.

Moreover, if

(iii) $\pi(\Ao)\subset \LL^\dagger(\D_\pi)$,

then $\pi$ is said to be a {  \it *-representation of the quasi
*-algebra} $(\A,\Ao)$.

\vspace*{3mm}

As for C*-algebras, we see here that a *-representation preserves
the algebraic structure of the abstract quasi *-algebra $(\A,\Ao)$.

{\bf Remark:--} It may be worth noticing that it might appear more
natural to represent $(\A,\Ao)$ in another quasi *-algebra
$(\Lc(\D_\pi,\D'_\pi),\Lc^\dagger(\D_\pi))$, analogous to the one
constructed above with $\D_\pi$ instead of $\D=D^\infty(N)$. We will
return on this quasi *-algebra in the following. Nevertheless, it is
usually more convenient to use $\Lc^\dagger(\D_\pi,\Hil_\pi)$ for
the following reasons:
\begin{enumerate}
\item if $a\in\A$ then $\pi(a)\in\Lc^\dagger(\D_\pi,\Hil_\pi)$. Therefore  $\forall \varphi\in\D_\pi$
 $\pi(a)\varphi\in\Hil_\pi$. Of
course, if we decide to represent $a$ as an element of
$\Lc(\D_\pi,\D'_\pi)$, we go out of the Hilbert space
($\pi(a)\varphi\notin\Hil_\pi$, in general)! This is not exactly
what one expects from a {\it representation} of a
*-algebra, since the abstract elements of the algebra are usually
represented acting and living on some Hilbert space;
\item we also have a technical reason to use
$\Lc^\dagger(\D_\pi,\Hil_\pi)$, which will appear clear in a moment:
in the theorem on the derivations given in the next section the
topology $\tau_s$ plays a role, and this can be defined on
$\Lc^\dagger(\D_\pi,\Hil_\pi)$ but not on $\Lc(\D_\pi,\D'_\pi)$;
\item finally we can observe that any partial *-algebra is a quasi
*-algebra: therefore the choice of representing a quasi *-algebra into a partial *-algebra
of operators is consistent.
\end{enumerate}

 The *-representation $\pi$ is called {  \it
ultra-cyclic} if there exists $\xi_0 \in \D_\pi$ such that
$\pi(\Ao)\xi_0=\D_\pi$. $\pi$ is {\it  faithful} if $\pi(x)=0$
implies $x=0$.

\vspace{1mm}

  As we have anticipated, we use $\pi$ to introduce a certain topology on $\pi(\A)$:
   let $\pi$ be a *-representation of $\A$.
{  \it The strong topology $\tau_s$} on $\pi(\A)$ is defined by the
 seminorms: $\{p_\xi(.); \; \xi\in\D_\pi\}$,
where $p_\xi(\pi(a))\equiv\|\pi(a)\xi\|$,  $a\in \A$, $\xi\in
\D_\pi$. This will be used in the next section. It may be worth
noticing that $\|\pi(a)\xi\|$ would make no sense in general if
$\pi(a)$ was an element of $\Lc^\dagger(\D,\D')$.

\vspace{2mm}

As for ordinary C*-algebras, even now it is possible to give a
{\it GNS-like construction}. As a matter of fact, several possible
extensions of this construction exist, but we will here mention
only one, \cite{camilloGNS}.

Let us assume here that the topology $\tau$ is given by a norm
$\|.\|$. Therefore $\A$ is a Banach space and we suppose, for
simplicity, that $(\A,\Ao)$ has a unit $\id$. Let $\varphi$ a
sesquilinear form on $\A\times\A$ such that

(i) $\varphi(x,x)\geq 0, \quad \forall x\in\A$;

(ii) $\varphi(ax,y)=\varphi(x,a^*y)$, $\forall a\in\A$, $x,y\in\Ao$;

(iii) there exists $\gamma>0$ such that $|\varphi(x,y)|\leq\gamma
\|x\|\,\|y\|$, $\forall x,y\in\Ao$.

\vspace{2mm}

 These conditions imply
that$$N_\varphi=\{a\in\A: \,\varphi(a,a)=0\} = \{a\in\A:
\,\varphi(a,b)=0,\,\forall b\in\A\}.$$ Let
$\lambda_\varphi(\A)=\A/N_\varphi$ and let us introduce a scalar
product on this vector space as follows:
$<\lambda_\varphi(a),\lambda_\varphi(b)>=\varphi(a,b)$.

Let further ${  \Hil_\varphi}$ be the completion of
$\lambda_\varphi(\A)$ wrt the norm inherited by this scalar
product. One can check that $\lambda_\varphi(\Ao)$ is dense in
$\Hil_\varphi$.

Let us finally define a map $\pi_\varphi^o$:
$$
\pi_\varphi^o(a)\lambda_\varphi(x)=\lambda_\varphi(ax), \quad
a\in\A, x\in\Ao:
$$
Then $\pi_\varphi^o$ is a *-representation of $\A$ in
$\Lc^\dagger(\lambda_\varphi(\Ao), \Hil_\varphi)$. Moreover,

(i) $\lambda_\varphi(\Ao)=\pi_\varphi^o(\Ao)\lambda_\varphi(\id)$
(i.e. $\lambda_\varphi(\id)$ is ultra-cyclic);

(ii)
$\varphi(a,b)=<\pi^o_\varphi(a)\lambda_\varphi(\id),\pi^o_\varphi(b)\lambda_\varphi(\id)>$,
$\forall a,b\in\A$.

\vspace{2mm}


{\bf   Remarks:--} (1)  this approach uses {\it   sesquilinear
forms} instead of {\it linear} functionals: indeed, while
$\omega(a^*b)$ is not well defined for generic $a,b\in\A$,
independently of the choice of the linear functional $\omega$,
$\varphi(a,b)$ surely makes sense if $\varphi$ is a generic
sesquilinear form on $\A\times\A$. This allows to define a scalar
product on $\A\times\A$, as shown above; \vspace{2mm}

(2) the {\it  physical interpretation} is analogous to that
discussed in the previous section: different sesquilinear forms
produce different representations which can still be interpreted as
different phases of the matter.

\section{Physical applications}

In this section we will show howthe algebraic framework discussed so
far can be of some usefulness in the rigorous treatment of some
physical systems.

\subsection{Existence of an effective hamiltonian}

We begin with reviewing some recent results obtained in
collaboration with A. Inoue and C. Trapani, \cite{bit1,bit2}, and
concerning the possibility of introducing, under certain
conditions on $\Sigma$, an effective hamiltonian.

  {\bf Definition:} Let
$(\A[\tau],\Ao)$ be a quasi *-algebra. A {\em *-derivation of }
$\Ao$ is a linear map $\delta: \Ao\rightarrow \A$
 with the following properties:
\begin{itemize}
\item[(i)]  $\delta(x^*)=\delta(x)^*, \; \forall x \in \Ao$;
\item [(ii)] $\delta(xy) = x\delta( y)+\delta( x)y,
\; \forall x,y \in \Ao$.
\end{itemize}
As we see, a *-derivation of $\Ao$ does exactly what one expects
from a similar object: it is a linear map, it preserves the adjoint,
and it satisfies the Leibnitz rule, of course only for those
elements for which this can be defined. From a physical point of
view we have already seen that it is quite hard for a derivation to
be {\it implemented} by an hamiltonian at an algebraic level: even
if a local hamiltonian does exist (i.e. the energy for finite $V$),
usually this sequence of operators do not converge to a self-adjoint
operator $H$ in most topologies, even if the sequence
$e^{iH_Vt}\,X\,e^{-iH_Vt}$ converges for each observable $X$. This
means that the dynamics is, in general, {\it hamiltonian} only at a
local level. However, as we have already discussed in Section III,
under some conditions of $\Sigma$ an {\it effective hamiltonian}
exists in $B(\Hil_\rho)$, i.e. in the C*-algebra obtained, via the
GNS-representation, from some state $\rho$. The role of the
representation appears evident now, and we will recover the
relevance of certain representations even in our settings. In
particular, we will restrict ourselves to those *-representations
$\pi$ of $(\A,\Ao)$ such that, whenever $x\in \Ao$ satisfies
$\pi(x)=0$, then $\pi(\delta(x))=0$. Under this rather natural
assumption, the linear map
$$
\delta_\pi(\pi(x))=\pi(\delta(x)), \quad x\in \Ao, $$is well-defined
on $\pi(\Ao)$ with values in $\pi(\A)$ and it is a
*-derivation of $\pi(\Ao)$. We call $\delta_\pi$ the {\it  *-derivation
 induced by $\pi$}.

Given such a representation $\pi$ and its dense domain
$\D_\pi\subset\Hil_\pi$, we consider the  graph topology $t_\dagger$
generated by the seminorms $$ \xi\in\D_\pi \rightarrow \|A\xi\|,
\quad A\in \LL^\dagger(\D_\pi). $$

{ Let $\D_\pi'$ be the conjugate dual space of $\D_\pi$ and
$t_\dagger'$  the {  strong dual topology} of $\D_\pi'$, i.e., see
our second appendix, the topology generated by the seminorms
$$
\D_\pi'\ni z\rightarrow \rho_{\cal E}(z):=\sup_{x\in {\cal
E}}|<x,z>|,
$$
where $<,>$ is the form which puts in duality $\D_\pi$ and
$\D_\pi'$ and $\cal E$ is a bounded set in $\D_\pi$. Then we get
the usual rigged Hilbert space
$$\D_\pi[t_\dagger] \subset \Hil_\pi \subset \D_\pi'[t_\dagger'].$$

Let $\LDD$ denote the space of all continuous linear maps from
$\D_\pi[t_\dagger]$ into $\D_\pi'[t_\dagger']$. Then one has
$$ \LL^\dagger(\D_\pi) \;  \subset
\; \LDD.$$}


\vspace*{2mm}

Each operator $A\in \LL^\dagger(\D_\pi)$ can be extended to an
operator $\hat A$ on the whole $\D_\pi'$ in the following way:
$$
<\hat A\xi',\eta>=<\xi',A^\dagger \eta>, \quad \forall \xi'\in
\D_\pi', \, \eta\in \D_\pi.
$$
Therefore the left and right multiplication of
$X\in\LL(\D_\pi,\D_\pi')$ and $A\in\LL^\dagger(\D_\pi)$ can always
be defined:
$$
(X\circ A)\xi=X(A\xi), \mbox{ and } (A\circ X)\xi=\hat A(X\xi), \,
\forall \xi\in \D_\pi,
$$
and for that we can conclude, as already anticipated before, that
{ $(\LL(\D_\pi,\D_\pi'),\LL^\dagger(\D_\pi))$ is a quasi
*-algebra.}

\vspace*{3mm}

Let $\delta$ be a *-derivation of $\Ao$ and $\pi$ an ultra-cyclic
*-representation of $(\A,\Ao)$ with ultra-cyclic vector $\xi_0$.
Then $\pi(\Ao)\subset\LL^\dagger(\D_\pi)$. We say that the
*-derivation $\delta_\pi$ induced by  $\pi$  is {  \it spatial} if
there exists $H_\pi=H_\pi^\dagger\in \LL(\D_\pi,\D_\pi')$ such
that $H_\pi\xi_0\in \Hil_\pi$ and
$$
\delta_\pi(\pi(x))=i\{H_\pi\circ\pi(x)-\pi(x)\circ H_\pi\},\quad
\forall x\in\Ao. $$

\vspace*{2mm} The meaning of this definition is clear: a derivation
produces in a representation $\pi$ a spatial induced derivation if
this can be {\it implemented by} a symmetric operator $H_\pi$: the
way in which $\delta_\pi$ is, in a certain obvious sense, described
by $H_\pi$ is via a generalized commutator, i.e. a commutator where
we may need to consider some adequate extensions of the the
operators involved. To be more explicit, if $x\in\Ao$, then
$\pi(x)\in\LL^\dagger(\D_\pi)$, so that, since by definition
$H_\pi\in \LL(\D_\pi,\D_\pi')$, it is clear that
$H_\pi\circ\pi(x)\psi=H_\pi\,\pi(x)\psi\in\D_\pi'$ for each
$\psi\in\D_\pi$. Viceversa, in general we have $\pi(x)\circ
H_\pi\psi=\hat\pi(x)H_\pi\psi\in\D_\pi'$, since $H_\pi\psi\in
\D_\pi'$ for each $\psi\in\D_\pi$, so that $\pi(x)(H_\pi\Psi)$ is
not well defined. This kind of difficulties, however, is not a big
surprise here since they are almost everywhere whenever one deals
with unbounded operators., and it is easily overcome here by means
of the $\circ$ multiplication.

\vspace*{2mm}

The main result concerning spatial derivations is contained in the
following theorem, \cite{bit1}, which extends and analogous result
for C*-algebras which can be found, for instance, in \cite{br}.

\begin{thm}

Let $(\A[\tau],\Ao)$ be a locally convex quasi *-algebra with
identity and $\delta$ be a *-derivation of $\Ao$.

Then the following statements are equivalent:

\vspace{3mm}

(i) There exists a $(\tau-\tau_s)$-continuous, ultra-cyclic
*-representation $\pi$ of $\A$, with ultra-cyclic vector $\xi_0$,
such that the *-derivation $\delta_\pi$ induced by  $\pi$  is
spatial.

\vspace{2mm}

(ii) There exists a positive linear functional $f$ on $\Ao$ such
that $ f(x^*x)\leq p(x)^2,$ $\forall x\in \Ao, $ for some continuous
seminorm $p$ of $\tau$ and, denoting with $\tilde f$ the continuous
extension of $f$ to  $\A$, the following inequality holds:
$$
|\tilde f(\delta(x))|\leq
C\left(\sqrt{f(x^*x)}+\sqrt{f(xx^*)}\right), \quad \forall x\in \Ao,
$$for some positive constant $C$.

\vspace*{4mm}

(iii) There exists a positive sesquilinear form $\varphi$ on
$\A\times\A$ such that:

$\varphi$ is invariant, i.e. $ \varphi(ax,y)=\varphi(x,a^*y),$ for
all $a\in \A$ and $x,y\in\Ao; $

$\varphi$ is $\tau$-continuous, i.e. $ |\varphi(a,b)|\leq p(a)
p(b),$ for all $a,b\in \A, $ for some continuous seminorm $p$ of
$\tau$;

$\varphi$ satisfies the following inequality: {\normalsize$$
|\varphi(\delta(x),\1)|\leq
C\left(\sqrt{\varphi(x,x)}+\sqrt{\varphi(x^*,x^*)}\right), \quad
\forall x\in \Ao, $$} for some positive constant $C$.
\label{theorem41}
\end{thm}

\vspace{3mm}

{\bf   Remarks:--} (1) even if $\delta$ cannot be written as
$\delta(x)=i[H,x]$, for any $H\in\A$, if the above theorem can be
applied, then $\delta_\pi$ is, essentially, the commutator with a
certain symmetric operator, $H_\pi$. Again, as we expect for
physical reasons, {\it the dynamics depends on the representation
$\pi$ and, as a consequence, on the phase of the matter}.

\vspace{2mm}

(2)  The above theorem can be used to answer to the following
question:  suppose we add to a spatial *-derivation $\delta_0$ a
{\it perturbation} $\delta_p$ such that $\delta=\delta_0+\delta_p$
is again a *-derivation. Under which conditions  is $\delta$ still
spatial? A sufficient condition for this to be true, \cite{bit1}, is
that $|\tilde f(\delta_p(x))|\leq |\tilde f(\delta_0(x))|$, for all
$x\in\Ao$, which is exactly what we expect since this means simply
that $\delta_p$ is {\it   smaller than } $\delta_0$. If we call
$H_\pi, H_{\pi,0}$ and $H_{\pi,p}$ the operators which implement
$\delta, \delta_0$ and $\delta_p$, we can also prove that
$i[H_\pi,A]\psi=i[H_{\pi,0}+H_{\pi,p},A]\psi$, for all
$A\in\Lc^\dagger(\D_\pi)$ and $\psi\in\D_\pi$.

\vspace{3mm} This theorem is the starting point to consider the
problem of the removal of the cutoff. This means that we are
assuming, as usual, that the dynamical behavior of the infinite
system is obtained as a suitable limit of its restriction to a
finite volume $V_L$. At the infinitesimal level, this means that we
have a family of inner derivations $\delta_L$, with $h_L$ their
associated energies, (i.e. $\delta_L(x)=i[h_L,x]$ for all $x\in\Ao$
and for all $L$) but we don't now if the limit of these derivations
is still inner or, at least, if the induced limiting derivation is
spatial in some particular representations. Let us now be more
precise.

Let $\Sys=\{(\A,\Ao),\Sigma,\alpha^t\}$ be a physical system, where,
extending Sewell's notation, \cite{sewbook2}, $(\A,\Ao)$ is a quasi
*-algebra, $\Sigma$ the set of states over $(\A,\Ao)$ and $\alpha^t$
the time evolution. Let further $\{\Sys_L=\{\A_L\subset
\Ao,\Sigma,\alpha_L^t\}, L\in\Lambda\}$ be a family of {\it
regularized} systems, i.e. $\Sys_L$ is the restriction of $\Sys$ to
some finite volume $V_L$. Here $\Lambda$ is a set of indexes, used
to label the finite volume systems $S_L$. We suppose here that, for
each fixed $L$, the dynamics $\alpha_L^t$ is generated by a
*-derivation $\delta_L$:
$$\alpha_L^t(x)=\tau_0-\sum_{k=0}^\infty\,\frac{t^k}{k!}\,\delta_L^k(x)=e^{ih_Lt}xe^{-ih_Lt},\quad\forall
x\in\A_L.$$ Here $\tau_0$ is a topology on $\Ao$. Actually, this
assumption is not necessary even if this is really what we have in
mind, and what actually happens for ordinary C*-algebras at least
for discrete systems, where each $h_L$ is a bounded operator.

\begin{defn}
The family $\{\Sys_L, L\in\Lambda\}$  is said to be {
c-representable} if there exists a *-representation $\pi$ of
$(\A,\Ao)$ into some $(\LL(\D_\pi,\D_\pi'),\LL^\dagger(\D_\pi))$
such that:

(i)  $\pi$ is $(\tau-\tau_s)$-continuous;

(ii) $\pi$ is ultra-cyclic with ultra-cyclic vector $\xi_0$;

(iii) when $\pi(x)=0$, then $\pi(\delta_L(x))=0$, $\forall
L\in\Lambda$.

\noindent Any such representation $\pi$ is a {
c-representation}.

\end{defn}

Making use of this definition we can prove the following
Proposition, \cite{bit2}:

\begin{prop}

Let $\{\Sys_L, L\in\Lambda\}$ be a c-representable family and
$\pi$ a c-representation. Let $h_L=h_L^*\in\A_L$ be the element
which implements $\delta_L$: $\delta_L(x)=i[h_L,x]$,  $\forall
x\in\Ao, \,\forall L\in\Lambda$. Suppose that $\delta_L(x)$ is
$\tau$-Cauchy $\forall x\in\Ao$ and that
$\sup_L\|\pi(h_L)\xi_0\|<\infty$.

Then, one has

{  (a) $\delta(x)=\tau-\lim_L\delta_L(x)$ exists in $\A$ and is a
*-derivation of $\Ao$;

(b) $\delta_\pi$, the
*-derivation induced by $\pi$, is well defined and spatial.}
\end{prop}

\vspace*{2mm}

{\bf   Remarks:--} (1) It is clear that if the sequence $\{h_L\}$ is
$\tau$-convergent, then $\delta_L(x)$ is automatically $\tau$-Cauchy
$\forall x\in\Ao$.

     \vspace*{2mm}

(2) This Proposition implies that any physical system $\Sys$ with
a c-representable regularized family $\{\Sys_L, L\in\Lambda\}$
{\it admits an effective hamiltonian in the sense of }
\cite{bt3,thi,sewbook1,lass}.  \vspace*{2mm}

(3) This is our version of Sewell's result  on the existence of
different {\it  effective hamiltonians} in different (GNS-like)
representations: a physical system $\Sigma$ exhibits different
dynamics in its different thermodynamical phases. \vspace*{2mm}

(4) We want to cite here an open problem which, in our opinion,
deserves a deeper analysis: what  happens if we consider a
representation $\pi'$ {\it
  globally equivalent} but locally different from a given c-representation $\pi$? Is $\pi'$ still a
c-representation? Do the related effective hamiltonians coincide?
This is indeed what one expects in connection with what has been
discussed in the C*-algebraic approach, even if no explicit proof of
this claim exists at this stage.

\subsection{The time evolution $\alpha^t$}

In this subsection we will consider the problem of the existence
of the algebraic dynamics for a system with infinite degrees of
freedom at the level of automorphisms of a certain quasi
*-algebra, instead of considering only its infinitesimal behavior.
More explicitly, the problem is the following: suppose that we have
been able to prove that the derivation $\delta$ exists.
Nevertheless, in general we have no information about $\delta^2$,
$\delta^3,\ldots$. Moreover, even if all these maps do exist, this
does not mean that the series
$\sum_{k=0}^\infty\,\frac{t^k}{k!}\,\delta^k(x)$, which defines
$\alpha^t(x)$ when $x$ and $H$ are both bounded, exists as well, for
a generic $x\in\Ao$. In other words, the existence of $\delta(x)$
does not imply, by no means, the existence of $\alpha^t(x)$ for
$x\in\A$ or even for $x\in\Ao$.

Furthermore, the effective hamiltonian $H_\pi$, whose existence
has been proved in the previous subsection, is  symmetric but not
self-adjoint, and therefore the spectral theorem cannot be used to
define $e^{iH_\pi t}$ and, as a consequence, to conclude that
$\pi(\alpha^t(x))=e^{iH_\pi t}\,\pi(x)\,e^{-iH_\pi t}$! So the
following crucial problem arises:  how to define a {\it time
evolution} in this case?

This is a rather hard problem already in a standard setting, when
there is no problem in multiplying elements of the algebra. Here
this is quite a dangerous operation and the difficulties are even
more than before. We will devote the rest of this subsection to
some partial results which can be used to analyze certain classes
of physical systems. It may be stressed that no general result
really exists at this stage.

\subsubsection{  From $\delta$ to $\alpha^t$: the first way}

We begin by considering a class of models which are suggested by the
{\it  mean field spin models}, reviewing some results first obtained
in \cite{bit2}.

 Let us assume that the finite volume hamiltonians $h_L$ can be
written in terms of some s.a. elements $s_L^\alpha$,
$\alpha=1,2,..,N$, which are $\tau$-converging to some elements
$s^\alpha\in\A$, commuting with all the elements of $\Ao$:
\vspace{-2mm}$$ s^\alpha=\tau-\lim_Ls_L^\alpha, \hspace{1cm}
[s^\alpha,x]=0,\, \forall x\in\Ao.$$ For mean field spin models
$s^\alpha$ is the {\it magnetization} and $\tau$ is the strong
topology restricted to a relevant family of states, \cite{bagmor},
or, alternatively, the so-called physical topology,
\cite{bt2,bt3,bt4,lass}, adopting the Lassner's terminology.

We say that $\{s_L^\alpha\}$  is  {\it  uniformly
$\tau$-continuous} if, for each continuous seminorm $p$ of $\tau$
and for all $\alpha=1,2,...,N$, there exists another continuous
seminorm $q$ of $\tau$ and a positive constant $c_{p,q,\alpha}$
such that $$ p(s_L^\alpha a)\leq c_{p,q,\alpha}q(a), \, \forall
a\in\A, \,\forall L\in\Lambda.$$

From this definition it also follows that $p(as_L^\alpha)\leq
c_{p,q,\alpha}q(a)$, $\forall a\in\A$, and that the same
inequalities can be extended to $s^\alpha$. Then we have

\begin{lemma}
If $\{s_L^\alpha\}$  is a uniformly $\tau$-continuous sequence and
if $\tau-\lim_Ls_L^\alpha=s^\alpha$, $\forall\alpha$, then {
$\tau-\lim_L\left(s_L^\alpha\right)^k=\left(s^\alpha\right)^k$},
$\forall\alpha$ and for $k=1,2,...$.\vspace{-2mm}
\end{lemma}

\begin{prop}
 Suppose that
(1) $\forall x\in\Ao$ $[h_L,x]$ depends on $L$ only through
$s_L^\alpha$ and (2) $s_L^\alpha\stackrel{\tau}{\longrightarrow}
s^\alpha$ and $\{s_L^\alpha\}$  is a uniformly $\tau$-continuous
sequence.

 Then, { for each $k\in \mathbb{N}$,}  the following
limit exists\vspace{-3mm}
$$
\tau-\lim_Li^k[h_L,x]_k=\tau-\lim_L\delta_L^k(x), \, \forall
x\in\Ao,\vspace{-3mm}$$ and defines an element of $\A$ which we call
{  $\delta^{(k)}(x)$}.
\end{prop}

\vspace{2mm}

{\bf   Remark:--}   The reason why we prefer to use
$\delta^{(k)}(x)$ instead of $\delta^k(x)$ is just to stress in this
way that it is not possible to write $\delta^k(x)=i^k[h,x]_k$, since
first of all no global $h$ does exist and, secondly, even if it
does, $[h,x]_k$ is not well defined in general because of domain
difficulties.

\vspace{2mm}

Once we have obtained conditions for all the multiple commutators
to exist in some reasonable sense, we still need to find
conditions for which the infinite series which defines
$\alpha^t(x)$ do converge. For that it is convenient to introduce
here the following definition:

\begin{defn}
We say that {\it  $x\in \Ao$ is a generalized analytic element of
$\delta$} if, for all $t$, the series
$\sum_{k=0}^\infty\frac{t^k}{k!}\pi(\delta^{(k)}(x))$ is
$\tau_s$-convergent. The set of all generalized analytic elements is
denoted with ${\cal G}$. \label{Definition 43}
\end{defn}

Therefore we have,\cite{bit2},

\begin{prop}
Let $x_\gamma$ be a net of elements of $\Ao$ and suppose that,
whenever $\pi(x_\gamma)\stackrel{\tau_s}{\longrightarrow}\pi(x)$
then $x_\gamma\stackrel{\tau}{\rightarrow}x$. Then, $\forall x\in
{\cal G}$ and $\forall t\in \mathbb{R}$, the series
$\sum_{k=0}^\infty\frac{t^k}{k!}\delta^{(k)}(x)$ converges in the
$\tau$-topology to an element of $\A$ which we call $\alpha^t(x)$.

Moreover, $\alpha^t$ can be extended to the $\tau$-closure
$\overline{\cal G}$ of ${\cal G}$. \label{theorem42}
\end{prop}

It may be worth noticing that, even if the assumptions are rather
strong, they are satisfied, for instance, by mean field spin
models!

\subsubsection{  From $\delta$ to $\alpha^t$: the second way}

Let $\pi$  be a faithful *-representation of the quasi *-algebra
$(\A,\Ao)$ and $\delta$  a *-derivation of $\Ao$ such that
$\delta_\pi$, is well-defined on $\pi(\Ao)$ with values in
$\pi(\A)$.

We define the following subset of $\Ao$ (a {\it  domain of
regularity } of $\delta$) $$
\Ao(\delta):=\{x\in\Ao:\quad\delta^k(x)\in\Ao,\quad \forall
k\in\mathbb{N}_0\}.$$ Whenever $\delta$  is {\it  regular} the set
$\Ao(\delta)$ is large. For instance, if $\delta$ is inner in
$\Ao$ with an implementing element $h\in\Ao$, then
$\Ao(\delta)=\Ao$. For general $\delta$, $\Ao(\delta)$ contains,
at least, all the multiples of the identity $\1$ of $\Ao$.

    \vspace{2mm}

$\Ao(\delta)$ is a *-algebra which is mapped into itself by
$\delta$. Moreover it is easy to check that
$\pi(\delta^k(x))=\delta_\pi^k(\pi(x))$, $\forall \,x\in\Ao(\delta)$
and  $\forall\,k\in\mathbb{N}_0$. Therefore it follows that
$\delta_\pi^k(\pi(x))\in\pi(\Ao)$.

{Let $\sigma_s$ be the topology on $\A$ defined via $\tau_s$ in the
following way:} $$\A\ni a\rightarrow
q_\xi(a)=p_\xi(\pi(a))=\|\pi(a)\xi\|, \quad \xi\in\D_\pi.
$$Then we have the following theorem, \cite{bit2}:

\begin{thm}

Let $(\A,\Ao)$ be a quasi *-algebra with identity, $\delta$ a
*-derivation on $\Ao$ and $\pi$ a faithful *-representation of
$(\A,\Ao)$ such that the induced derivation $\delta_\pi$ is well
defined. Then, we have:    \vspace{2mm}

(1) if the following inequality holds {\normalsize$$ \forall
\eta\in\D_\pi \, \exists c_\eta>0: p_\eta(\delta_\pi(\pi(x)))\leq
c_\eta p_\eta(\pi(x)), \quad \forall x\in \Ao(\delta), $$} then
$\sum_{k=0}^\infty\frac{t^k}{k!}\delta^k(x)$ converges for all $t$
in the topology $\sigma_s$ to an element of
$\overline{\Ao(\delta)}^{\sigma_s}$ which we call $\alpha^t(x)$;
$\alpha^t$ can be extended to $\overline{\Ao(\delta)}^{\sigma_s}$.
Moreover $\alpha^t:\overline{\Ao(\delta)}^{\sigma_s}\rightarrow
\overline{\Ao(\delta)}^{\sigma_s}$ and
$$\alpha^{t+\tau}(x)=\alpha^t(\alpha^\tau(x)), \quad \forall t, \tau,
\, \forall x\in\Ao(\delta);$$

\vspace*{1mm}

(2) Suppose that the following inequality holds
{\normalsize$$\exists c>0: \, \forall \eta_1\in\D_\pi \,\, \exists
A_{\eta_1}>0,\, n\in\mathbb{N} \mbox{ and } \eta_2\in\D_\pi:$$
$$p_{\eta_1}(\delta_\pi^k(\pi(x)))\leq A_{\eta_1} c^k k! k^n
p_{\eta_2}(\pi(x)), \quad \forall x\in \Ao(\delta),\,\forall
k\in\mathbb{N}_0, $$}\vspace{3mm}  then
$\sum_{k=0}^\infty\frac{t^k}{k!}\delta^k(x)$ converges, for
$t<\frac{1}{c}$ in the topology $\sigma_s$ to an element of
$\overline{\Ao(\delta)}^{\sigma_s}$ which we call $\alpha^t(x)$;
$\alpha^t$ can be extended to $\overline{\Ao(\delta)}^{\sigma_s}$.
Moreover $\alpha^t$ maps $\overline{\Ao(\delta)}^{\sigma_s}$ into
itself for $t<\frac{1}{c}$ and, $\forall x\in\Ao(\delta)$,
$$ \alpha^{t+\tau}(x)=\alpha^t(\alpha^\tau(x)), \quad \forall t,
\tau, \mbox{ with } t+\tau<\frac{1}{c}.$$

 \label{theoremextra1}
\end{thm}

     \vspace{2mm}

{\bf   Remarks:--} (1) As we see, this theorem gives sufficient
conditions for $\alpha^t$ to be defined (as a converging series)
at least on a certain subset of $\Ao$.

\vspace{2mm}

(2) Here and in the previous approach the spatiality of the
derivation is not required. It is obvious that, when $H_\pi$
exists as a {\bf self-adjoint} operator mapping $\D_\pi$ into
$\Hil_\pi$, we could use the spectral theorem to define
$\pi(\alpha^t(x))=e^{iH_\pi t}\pi(x)e^{-iH_\pi t}$;


\subsubsection{  A different point of view }

In a recent paper, \cite{bt4}, we have considered the problem of
the existence of $\alpha^t$ from a slightly different point of
view, which is maybe more suitable for systems with a finite
number of degrees of freedom. This is because we have assumed that
the energy operator of our quantum system system does exist as a
self-adjoint, unbounded and densely defined operator $H_0\geq \1$.
Then, it is known that the operator $e^{iH_0t}$, and therefore the
time evolution of an observable $X$, can be defined via the
spectral theorem. However, but for finite dimensional Hilbert
spaces, our claim is that the {\it natural algebraic framework} to
discuss the dynamical behavior of the system is $\LD[\tau_0]$,
where $\D=D^\infty(H_0)$, rather than $B(\Hil)$. Indeed, if
$dim(\Hil)=\infty$, it is clear that in general $H_0\notin
B(\Hil)$ and that $\delta$ does not map $B(\Hil)$ into itself. On
the other hand, it is evident that $H_0\in\LD[\tau_0]$ and that
$\delta:\LD[\tau_0]\rightarrow\LD[\tau_0]$.

These claims are based on the following natural procedure:

let $H_0 = \int_1^\infty \lambda d E(\lambda)$ be the spectral
decomposition of $H_0$, see Appendix 2. We define, for $L\geq 1$,
the projectors $ Q^0_L= \int_1^L d E(\lambda) $ and we introduce the
{\it regularized hamiltonian} $H_L = Q^0_L H_0 Q^0_L .$

For each $L$, we see that {  $Q^0_L, H_L\in B(\Hil)\bigcap \LD$}.
Furthermore, we have $[Q^0_L, H_L]=[Q^0_L, H_0]=[H_0, H_L]=0$.

If $\tau_0$ is the topology on $\LD$ generated by the seminorms
$$\LD \ni A \mapsto \|A\|^{f,k} = \max\{
\|H_0^kAf(H_0)\|,\|f(H_0)AH_0^k\|\},$$ then we have:

\vspace*{3mm}

{  (i)} $H_L \to H_0$ with respect to the topology $\tau_0$;

    \vspace{2mm}

{  (ii)} $\left\{{\rm e}^{itH_L}\right\}$ is $\tau_0$-Cauchy in
$\LD$ and converges to $e^{iH_0t}$

    \vspace{2mm}

{  (iii)}  $\forall A \in \LD$, the sequence $\left\{{\rm
e}^{itH_L}A{\rm e}^{-itH_L}\right\}$ is $\tau_0$-Cauchy in $\LD$
and converges to ${\rm e}^{itH_0}A{\rm e}^{-itH_0}$ .

    \vspace{3mm}

We can therefore conclude that  $H_0,\, e^{iH_0t},$ and
$\alpha^t(A):={\rm e}^{itH_0}A{\rm e}^{-itH_0}$ all belong to
$\LD$, $\forall A\in\LD$. Moreover we can also show that
$$\alpha^t(A)=\tau_0-\lim_L {\rm e}^{itH_L}A{\rm
e}^{-itH_L}= \left(\tau_0-\lim_L {\rm
e}^{itH_L}\right)A\left(\tau_0-\lim_L {\rm e}^{-itH_L}\right).$$

This suggests the use of $\LD[\tau_0]$ as a natural algebraic and
topological framework for the analysis of the time evolution of, at
least, finite quantum systems. Of course, a similar construction can
be repeated also for $QM_\infty$ systems, at least for those systems
for which an unbounded, self-adjoint and densely defined operator
$M$ exists such that $[M,H_L]=0$ (on a dense domain), \cite{lass}.

 In the same paper we have
considered the role of a perturbation in this approach: let
$H=H_0+B$, and suppose that the spectral decomposition of the {\it
free hamiltonian} $H_0$ is explicitly known while the spectral
decomposition of the {\it perturbed hamiltonian} $H$ cannot be
exactly found, which is exactly what usually happens in concrete
situations. We have shown that the {\it convenient algebraic
structure} is again $\LD$, with $\D=D^\infty(H_0)$, (since, if $H_0$
has discrete spectrum, we know an o.n. set in $\D$ and, as a
consequence, we know $\D$) but the { {\it technically convenient}
topology}, $\tau$, is that given by the seminorms $$\LD \ni A
\mapsto \|A\|_+^{f,k} = \max\{ \|H^kAf(H)\|,\|f(H)AH^k\|\},$$because
with this choice some of the above convergence results can be
established. Moreover this apparent difference between the algebraic
and the topological frameworks, can be easily controlled. Indeed we
have proven in \cite{bt4} that, if {(1)} $D(H_0)\subseteq D(B)$ and
if $H=H_0+B$ is self-adjoint on $D(H_0)$, and {(2)} if
$\D^\infty(H_0) = \D^\infty(H)$ (hypothesis for which we gave
necessary and sufficient conditions), then $\tau_0\equiv\tau$.

Under these assumptions we can therefore undertake a deeper
analysis of the existence of the algebraic dynamics for a
perturbed hamiltonian. We refer to \cite{bt4} for a detailed
analysis, which eventually produces a rigorous definition of the
Schr\"odinger dynamics.

\subsection{Fixed point results}

This is an alternative procedure which again produces a rigorous
definition of the dynamics of a (closed) physical system,
\cite{bag2}, and which is based on a generalization of well known
fixed point theorems.

\vspace{2mm}

Let $\B$ be a $\tau$-complete subspace of $\Lc^\dagger(\D)$ and $T$
a map from $\B$ into $\B$. We say that $T$  is a {\em weak $\tau$
strict contraction over $\B$}, briefly a w$\tau$sc$(\B)$, if there
exists a constant $c \in]0,1[$ such that, for all
$(h,k)\in\C_N:=(\C,{\Bbb{N}}_0)$, ${\Bbb{N}}_0={\Bbb{N}}\cup\{0\}$,
there exists a pair $(h',k')\in\C_N$ satisfying \be
\|Tx-Ty\|^{h,k}\leq c \|x-y\|^{h',k'}\hspace{2cm}\forall \: x,y \in
\B.  \label{31} \en

In what follows we will consider equations of the form {$Tx=x$},
$T$ being a w$\tau$sc($\B$). The first step consists in
introducing the following subset  of $\B$: \be
\B_L\equiv\left\{x\in \B: \sup_{(h,k)\in\C_N}\|Tx-x\|^{h,k}\leq
L\right\}, \label{313} \en $L$ being a fixed positive real number.

\begin{lemma} {Let $T$ be a w$\tau$sc($\B$). Then

(a) if $T0=0$ then any $x\in \B$ such that $\sup_{(h,k)\in\C_N}\|
x\|^{h,k}\leq L_1$ belongs to $\B_L$ for $L\geq L_1(1+c)$;

(b) if $\|T0\|^{h,k}\leq L_2$ for all $(h,k)\in\C_N$, then any
$x\in \B$ such that $\sup_{(h,k)\in\C_N}\| x\|^{h,k}\leq L_1$
belongs to $\B_L$ for $L\geq L_1(1+c)+L_2$;

(c) if $x\in \B_L$ then $T^n x\in \B_L$, for all $n\in \N$;

(d) $\B_L$ is $\tau$-complete;

(e) if $\B_L$ is not empty, then $T$ is a w$\tau$sc($\B_L$).}

\end{lemma}

\vspace{2mm}

$\B_L$ is non empty, see \cite{bag2}. The existence of a fixed
point is ensured by the following Proposition:

\begin{prop}{ Let $T$ be a w$\tau$sc($\B$). Then

(a) $\forall x_0\in \B_L$ the sequence $\{x_n\equiv
T^nx_0\}_{n\geq 0}$
 is $\tau$-Cauchy in $\B_L$. Its $\tau$-limit, $x\in \B_L$, is a fixed
point of $T$;

(b) if $x_0,y_0\in \B_L$ satisfy the condition
$\sup_{(h,k)\in\C_N}\|x_0-y_0\|^{h,k}<\infty,$ then
$\tau-\lim_nT^nx_0=\tau-\lim_nT^ny_0$.}

\end{prop}

\vspace{.6cm}

For physical applications  we need to consider the case in which
these maps depend on an external parameter:

let $I\subset \R$ be a set such that $0$ is one of its accumulation
points. A family of weak $\tau$ strict contractions
$\{T_\alpha\}_{\alpha\in I}$ is said to be {\em uniform} if

1) $T_\alpha:\B \rightarrow \B \:\: \forall \alpha\in I$, $\B$
being a $\tau$-complete subspace of $\Lc^+(\D)$;

2) $\forall (h,k)\in \C_N$ and $\forall \alpha \in I$ there exist
$(h',k') \in \C_N$,  independent of $\alpha$, and  $c_\alpha\in
]0,1[$, independent of $(h,k)$, such that \be \|T_\alpha x-
T_\alpha y\|^{h,k}\leq c_\alpha \|x- y\|^{h',k'}, \quad \forall
x,y \in \B; \label{42fp} \en

3) $c_-\equiv \lim_{\alpha, 0}c_\alpha \in]0,1[$.

\vspace{2mm}

We further say that the family $\{T_\alpha\}_{\alpha\in I}$ is
{$\tau$-strong Cauchy} if, for all $(h,k)\in \C_N$ and $\forall
y\in\B$,
 \be
\|T_\alpha y- T_\beta y\|^{h,k}\stackrel{\alpha,\beta\rightarrow
0}{\longrightarrow} 0. \quad \label{44} \en

We call $\B_L^{(\alpha)}$ the following set
$\B_L^{(\alpha)}\equiv\left\{x\in \B:
\sup_{(h,k)\in\C_N}\|T_\alpha x-x\|^{h,k}\leq L\right\}.$

\vspace{3mm}

\begin{prop}{Let $\{T_\alpha\}_{\alpha\in I}$ be a $\tau$-strong Cauchy
uniform family of w$\tau$sc($\B$). Then

(1) There exists a w$\tau$sc($\B$), $T$, which satisfies the
following relations: $$ \|T y- T_\alpha y\|^{h,k}\rightarrow 0
\quad \forall y\in\B, \, \forall (h,k)\in \C_N $$ and $$ \|T y- T
z\|^{h,k}\leq c_-\| y- z\|^{h',k'}\quad \forall y,z\in\B, $$ where
$( h',k')$ are those of inequality (\ref{42fp}).

(2) let $\{x_\alpha\}_{\alpha\in I}$ be a family of fixed points
of the net $\{T_\alpha\}_{\alpha\in I}$: $ T_\alpha
x_\alpha=x_\alpha$, $\forall \alpha\in I$. If
$\{x_\alpha\}_{\alpha\in I}$ is a $\tau$-Cauchy net then, calling
$x$ its $\tau$-limit in $\B$,  $x$ is a fixed point of $T$.

(3) If the set $\cap_{\alpha\in I} \B_L^{(\alpha)}$ is not empty
and if the following commutation rule holds

$$ T_\alpha (T_\beta y) = T_\beta (T_\alpha y), \quad \forall
\alpha,\beta\in I  \mbox{ and } \forall y\in \B, $$
 then, calling
$$ x_\alpha=\tau-\lim_{n\rightarrow \infty} T_\alpha^nx^0,
\mbox{where } x^0\in \cap_{\alpha\in I} \B_L^{(\alpha)}, $$ each
$x_\alpha$ is a fixed point of  $T_\alpha$,  $T_\alpha
x_\alpha=x_\alpha $ and $\{x_\alpha\}_{\alpha\in I}$ is a
$\tau$-Cauchy net. Moreover $\tau-\lim_{\alpha\rightarrow 0}
x_\alpha$ is a fixed point of $T$.}

\end{prop}

\vspace{.4cm}

As an application we have proven in \cite{bag2} that, under
certain technical assumptions, the time evolution of a given
operator $x$,
$$ x_\alpha(t)=x+i\int_0^tds [H_\alpha,x_\alpha(s)], $$ is associated with a uniform family of
w$\tau$sc($\Lc^+$), $\{U_\alpha\}$, which is also $\tau$-strong
Cauchy. This implies that, because of the Proposition above, the
dynamics for the physical system can be obtained as a $\tau$-limit
of the regularized dynamics $x_\alpha(t)$, which is a fixed point
of $U:=\lim_{\alpha} U_\alpha$.

\subsection{Explicit estimates}

We end this excursus of (class of) models for which the time
evolution is under control, by considering the so-called {almost
mean field Ising model}, defined by the following finite volume
hamiltonian
\begin{equation}
 H_V =
\frac{J}{|V|^\gamma}\sum_{i,j \in V}   \sigma^3_i \sigma^3_j,
\label{31bis}
\end{equation}
with $0<\gamma \leq 1$, \cite{bt1}. Particularly relevant in the
mathematical description of this model is the {\em almost
magnetization} operator $S_V^3:=\frac{1}{|V|^\gamma}\sum_{p\in
V}\sigma_p^3.$ In fact, if $A$ is a local observable, its
regularized time evolution $\alpha_V^t(A):=e^{iH_Vt}Ae^{-iH_Vt}$
in general depends  on $t$, $A$ and $S_V^3$.

In the first appendix it is discussed in some details how to
construct the relevant Hilbert space for the model, $\Hil_{\{n\}}$,
the dense domain $\D_{\{n\}}$, the O*-algebra $\Lc(\D_{\{n\}})$, a
*-representation of the C*-spin algebra $\A_s$ and the physical topology $\tau_0$,
following \cite{lass}. Here  we  introduce also a different topology
$\tau$ on $\A_s$, which has proved to be of some usefulness, as
follows:
$$
\tau:\hspace{9mm}\|A\|^{f}_{\{n\}}:=\|f(M_{\{n\}})\pi_{\{n\}}(A)f(M_{\{n\}})\|,
$$
 where
$f$ belongs to $\C$. With these definitions, calling $\Ao$ the
$\tau_0$-completion of $\A_s$ and $\A$ the $\tau$-completion of
$\A_s$, we proved in \cite{bt1} that:
\begin{itemize}
\item $(\A[\tau],\Ao[\tau_0])$ is a topological quasi *-algebra; \item
all the powers of the almost magnetization $S_3^V$ are
$\tau_0$-converging in $\A$; \item the finite volume dynamics
$\alpha_V^t$ $\tau_0$-converges to a one-parameter group of
automorphisms $\alpha^t$ of $\Ao$ ;\item $\alpha^t$ solves the
$\tau_0$-limit of the finite volume Heisenberg equation of motion.
\end{itemize}

\vspace{3mm}

Another spin model which can be analyzed within the same algebraic
framework is the {\em almost mean field Heisenberg model},
$$
 H_V =
\frac{J}{|V|^\gamma}\sum_{i,j \in V}  \sum_{\alpha=1}^3
\sigma^\alpha_i \sigma^\alpha_j,
$$
with $\frac{1}{2}<\gamma \leq 1$, see \cite{bt2}, which differs
from the Ising model because it is intrinsically
three-dimensional.

\vspace{2mm}

A different class of models that we have considered using the same
approach involves free and interacting bosons, \cite{bag1}.
 The formal hamiltonian $H$ for the one mode free
bosons is simply the number operator $N=a^\dagger a$, $a$ and
$a^\dagger$ being the annihilation and creation operators for the
bosons. They satisfy the canonical commutation relation
$[a,a^\dagger]=\I$. (More properly, $N$ is the unique self-adjoint
extension of the symmetric operator $a^\dagger a$.)

The construction of the topological quasi *-algebra is the usual
one. Let $ \D := D^\infty(N) = \cap_{k\geq 0} D(N^k). $ This set
is dense in the Fock-Hilbert space $\Hil$ constructed in the
standard way. Starting from $\D$ we can define the *-algebra
$\Lc^+(\D)$. It is clear that all powers of $a$ and $a^\dagger$
belong to this set.
 The topology in $\Lc^+(\D)$ is, using Lassner's terminology in \cite{lass}, the usual quasi-uniform topology:
\be X\in \Lc^+(\D) \rightarrow \|X\|^{f,k} :=
\max\left\{\|f(N)XN^k\|,\|N^kXf(N)\|\right\}, \label{42} \en where
$f\in \C$ and $k\geq0$. We have already discussed several times
along this paper that $\Lc^+(\D)[\tau_0]$ is a complete locally
convex topological *-algebra.

Let {$\E_l$} be the subspace of $\Hil$ generated by all the
vectors which are proportional to $(a^\dagger)^l\Phi_0$. Let also
{$\F_L$} be the direct sum $\F_L:=
\E_0\oplus\E_1\oplus.....\oplus\E_L$. Finally, let
$N=\sum_{l=0}^{\infty}l\Pi_l$ be the spectral decomposition of the
number operator $N$. The operators {$\Pi_l$} are projection
operators, as well as the operators {$Q_L=\sum_{l=0}^{L}\Pi_l$}.
The following properties are obvious:
$$ \Pi_k \Pi_l=\delta_{kl}\Pi_l, \hspace{.2cm}
\Pi_k^\dagger=\Pi_k;  \hspace{.5cm} Q_L Q_M=Q_L, \hspace{.2cm}
\mbox{if } L\leq M, \hspace{.5cm} Q_L^\dagger=Q_L.
$$
It is clear that {$\Pi_k:\Hil \rightarrow \E_k$}, and {$Q_L:\Hil
\rightarrow \F_L$}. The operator $Q_L$ is used to cut-off the
hamiltonian, by replacing $a$ with $a_L:=Q_LaQ_L$. The regularized
hamiltonian is simply $H_L=Q_LNQ_L=NQ_L$ and the related time
evolution is $\alpha_L^t(X)=e^{iH_Lt}Xe^{-iH_Lt}$. This {\em
occupation number cut-off} produces a self adjoint bounded
operator $H_L$ and we have shown in \cite{bag1} that
 {\em the limits of $\alpha_L^t(a^n)$
and $\alpha_L^t((a^\dagger)^n)$ exist in $\Lc^+(\D)[\tau_0]$   for
all $n\in \N$}. We have already seen that this result has been
generalized by those in \cite{bt4}.

\vspace{2mm}

The same algebraic framework turns out to be useful also in the
analysis of the thermodynamical limit of the interacting model
described by the following \underline{formal} hamiltonian:
$$
H_V=\frac{J}{|V|}\sum_{i,j\in V}\sigma^3_i\sigma^3_j+a^\dagger a
+\gamma (a+a^\dagger)\sigma^3_V,
$$
where $\sigma^3_V=\frac{1}{|V|}\sum_{i\in V}\sigma^3_i$. Here the
algebra $\Lc^+(\D)$ must be replaced by  $\A= B(\Hil_{spin})
\otimes \Lc^+(\D)$. The topology on $\A$, $\tau_{comp}$, is
generated by the following seminorms:$\|XA\|^{f,k,\Psi}\equiv
\|X\|^{f,k}
 \|A\Psi\|,$ $X\in \Lc^+(\D)$ and $A\in B(\Hil_{spin})$. It is
worthwhile to remind also that $\Psi$ cannot be a generic vector
in $\Hil_{spin}$, but must belong to the set  $$
 \F =\left\{ \Psi \in \Hil_{spin} \: :
\lim_{|V|,\infty} \frac{1}{|V|}\sum_{p\in V} {\sigma^3_p}\Psi =
\sigma^3_\infty \Psi, \:\: \|\sigma^3_\infty\|\leq1 \right\}.
$$
As before, the regularized hamiltonian is obtained by replacing
$a$ with $a_L:=Q_LaQ_L$, so that the new hamiltonian $H_{V,L}$
depends on two, in principle, unrelated cutoffs. The existence of
the limit of $\alpha_{V,L}^t(X)=e^{iH_{V,L}t}Xe^{-iH_{V,L}t}$ is
ensured by the following result, \cite{bag1}: {\em the limit of
$\alpha_{V,L}^t(a)$ for $|V|$ and $L$ both diverging exists in
$\A[\tau_0]$. Moreover, if the two cutoffs satisfy the relation
$|V|=L^r$, for a certain integer $r>1$, the same holds true also
for $\alpha_{V,L}^t(\sigma_\alpha^i)$}.

\subsection{Few words on other results}

In this paper we have only discussed in some details results related
to those research lines I am more involved which, as already
mentioned, are mainly related to quasi
*-algebras. We dedicate this short section to a very brief list
of different lines of research, starting with the analysis of {\em
one-parameter groups of *-automorphisms} in the context of a
particular class of partial *-algebras, the so-called {\em partial
O*-algebras}. This analysis is important because both time evolution
and physical symmetries are  examples of *-automorphisms. Some
results on the existence of the time evolution for a given physical
system, its continuity and the spatiality of the related derivation
can be found in \cite{aitbook} and in references therein.

Another application of algebras of unbounded operators originates
from the analysis of point-like quantum fields as discussed in
\cite{fred,jos65}. Here the field $A(x)$ is represented as a
sesquilinear form on a certain domain $\D\subseteq\Hil$. One of the
basic Wightman axioms is that the smeared field
$A(f)=\int_{{\Bbb{R}}^4}A(x)f(x)dx$ exists as a well defined
operator in $\D$ for any given $f\in\C^{\infty}_0({\Bbb{R}}^4)$.
However this may not be true and different possible definitions of
point-like field have been proposed in the literature. A detailed
analysis on this subject can be found, for instance, in \cite{epi}.

We end this short subsection mentioning a last application of quasi
*-algebras in the analysis of the dynamics of a free Bose system confined in a segment of length $l$.
A contradiction arising from the analysis of this system, which
originates from the use of the Bogoliubov inequality,  disappear
when one constructs the {\em CCR quasi
*-algebra} as in \cite{llt} or, alternatively, adopting the point of view of \cite{bou} where
the authors generalize the notion of states on unbounded operator
algebras.

Other physical applications of algebras of unbounded operators can
be found in \cite{dubhen}.

\section{Work in progress and future projects}

We want to discuss here some preliminary results concerning a
situation in which the algebraic framework is somehow fixed and no
{\it global} hamiltonian exists, but only a family of finite volume
energy operators. This is essentially what happens in the standard
formulation of $QM_\infty$. More in details, let $S$ be a
self-adjoint, unbounded, densely defined operator on a Hilbert space
$\Hil$. For simplicity we assume that its spectrum is discrete, even
if most of the results do not depend on this aspect:
$S=\sum_{l=0}^\infty s_lP_l$. Let $\D=D^\infty(S)$, and $\LD$ and
$\tau$ constructed as usual. Let further $H_L=\sum_{l=0}^L\,h_lP_l$
be our {\it   regular hamiltonian}: $H_L\in B(\Hil)$, $\forall L$.
It is worth stressing that we are assuming, for the time being, that
the spectral projections of $H_L$ are the same as those of $S$.

In some of our previous attempt, and in particular in what we have
discussed in Subsection V.2.3, we had $\tau-\lim H_L\in\LD$. This
implies, in particular, that $H$ exists and $H\in\LD$. In this case
we have
 seen that there is absolutely no problem in defining a
Shr\"odinger or an Heisenberg dynamics. In \cite{bagwin} we have
proven that this is not necessary. More in details, we have proven
that for each sequence $\{h_l\}$, if $\{s_l^{-1}\}$ is in
$l^2(\mathbb{N}_0)$, then

\begin{enumerate}

\vspace{-1mm}

\item $e^{iH_Lt}$ $\tau$-converges to an element $T_t\in\LD$;

\vspace{-1mm}

\item $\forall X\in\LD$ the sequence $e^{iH_Lt}Xe^{-iH_Lt}$
$\tau$-converges to an element $\alpha^t(X)\in\LD$;

\vspace{-1mm}

\item $\forall X\in\LD$ we have $\alpha^t(X)=T_tXT_{-t}$;

\item if $Q_M=\sum_{l=0}^M$, $X\in\LD$, $X_M=Q_MXQ_M$ and
$\delta_L(X_M)=i[H_L,X_M]$ then
$$\alpha^t(X)=\tau-\lim_{L,M,N}\,\sum_{j=0}^N\,\frac{t^j}{j!}\,\delta_L^j(X_M).$$

\end{enumerate}

\vspace{3mm}

{\bf  Remark:--} We see, therefore, that the time evolution of each
element of $\LD$ can be defined (in three different ways!) even if
$H_L$ does not define any {  hamiltonian} of the system $\Sigma$,
i.e. even if $H_L$ does not converge in {\bf  any} {\it natural
topology}. This is relevant for us since it is exactly what happens
in the most general physical situations, as we have widely discussed
in Section II. \vspace{2mm}

 However, here we are assuming that $S$ and $H_L$ admit the
same spectral projections. It is natural to ask  what happens if
this is no longer true. In this case we have the following partial
results, \cite{bagwin}:

\vspace*{2mm}

suppose that $S=\sum_{l=0}^\infty s_lP_l$ and $H_M=\sum_{l=0}^M
h_lE_l$, with $E_j\neq P_j$. Then something can be said also in this
case. In particular

{  $\bullet$} if $[E_l,P_j]=0$ for all $l,j$, or if $E_l\neq P_l$
only for a finite number of $l$'s, the above results still can be
proved;    \vspace{2mm}

{  $\bullet$} let $\{\varphi_l\}$ and $\{\psi_l\}$ be two different
orthonormal bases of $\Hil$ and suppose that
$P_l=|\varphi_l><\varphi_l|$ and $E_l=|\psi_l><\psi_l|$. It is clear
that $[E_l,P_j]\neq0$ in general. Nevertheless, if $\psi_l$ is a
finite linear combination of the $\varphi_j$'s, then again the above
results still can be proved. \vspace{2mm}

\vspace{3mm}

The last result of \cite{bagwin} which we want to cite here concerns
the role of the Gibbs and the KMS states for this situation: let
$\rho_L:=\frac{e^{-\beta H_L}}{tr_L\left(e^{-\beta H_L}\right)}$ be
the density matrix of a Gibbs state at the inverse temperature
$\beta$. Then it is easy to check that $\tau-\lim_L\rho_L$ exists in
$\LD$. But it is still to be investigated is whether this limit is a
KMS state (in some sense).


As it is clear even if many results have been obtained within this
context, many others are still to be obtained. In particular, the
following research lines are already opened:
\begin{enumerate}

\item we need a deeper analysis of the previous results when $S$ and
$H_L$ are \underline{essentially} different and, in particular, if
$[H_{L_1},H_{L_2}]\neq 0$.

\item What can be said  about Goldstone's theorem when $\alpha_V^t$ does not converge
uniformly (or $\cal F$-strongly) to $\alpha^t$? What does this
theorem become in a quasi *-algebraic framework?

\item Can we define a KMS state when $\alpha_V^t$ does not converge
uniformly (or $\cal F$-strongly) to $\alpha^t$?
\item Is there any relation between these {\it KMS-like states} and the  phase structure of the physical
system?

\item Is there any relation between these {\it KMS-like states} and the Tomita-Takesaki
theory? (something is discussed in \cite{aitbook})
\item What about local modifications? Do two states $\rho$ and $\chi$ which are only {\it locally
different} generate unitarily equivalent representations? And what
can be said about the related effective hamiltonians? (Some results
are already discussed in \cite{bt3})

\end{enumerate}

\bigskip
\noindent {\large\bf Acknowledgement}

This work has been financially supported in part by M.U.R.S.T.,
within the  project {\em Problemi Matematici Non Lineari di
Propagazione e Stabilit\`a nei Modelli del Continuo}, coordinated by
Prof. T. Ruggeri.

It is a pleasure to use this occasion to thank some friends that,
during these years, shared with me the interest in the algebras of
operators: C. Trapani, A. Inoue, G.L. Sewell, G. Morchio, F.
Strocchi and J.-P. Antoine, among  others. Also, I like to thank A.
Greco and T. Ruggeri because their invitation to the 2006 GNFM
meeting was the starting point for this review!




\newpage

\appendix
\renewcommand{\theequation}{\Alph{section}.\arabic{equation}}


 \section{\hspace{-14.5mm} Appendix 1:  the algebras for $\Sigma$}

Let $\Sigma$ be a system with infinite degrees of freedom. We recall
that the Haag ang Kastler's construction of  the C*-algebra
associated to $\Sigma$, can be schematized as follows:

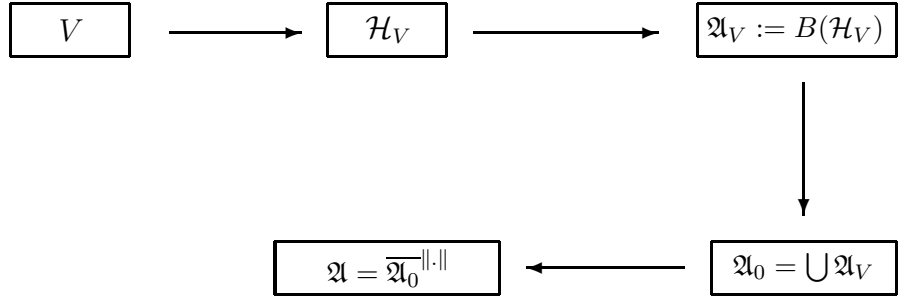
\begin{picture}(450,90)

\put(60,35){\thicklines\line(1,0){45}}
\put(60,55){\thicklines\line(1,0){45}}
\put(60,35){\thicklines\line(0,1){20}}
\put(105,35){\thicklines\line(0,1){20}}
\put(83,45){\makebox(0,0){$V$}}

\put(120,44){\thicklines\vector(1,0){50}}

\put(180,35){\thicklines\line(1,0){45}}
\put(180,55){\thicklines\line(1,0){45}}
\put(180,35){\thicklines\line(0,1){20}}
\put(225,35){\thicklines\line(0,1){20}}
\put(203,45){\makebox(0,0){$\Hil_V$}}

\put(235,44){\thicklines\vector(1,0){70}}

\put(320,35){\thicklines\line(1,0){75}}
\put(320,55){\thicklines\line(1,0){75}}
\put(320,35){\thicklines\line(0,1){20}}
\put(395,35){\thicklines\line(0,1){20}}
\put(357,45){\makebox(0,0){\small$\A_V:=B(\Hil_V)$}}

\put(360,25){\thicklines\vector(0,-1){50}} %

\put(160,-55){\thicklines\line(1,0){85}}
\put(160,-35){\thicklines\line(1,0){85}}
\put(160,-55){\thicklines\line(0,1){20}}
\put(245,-55){\thicklines\line(0,1){20}}
\put(203,-45){\makebox(0,0){\small{$\A=\overline{\Ao}^{\|.\|}$}}}

\put(315,-45){\thicklines\vector(-1,0){60}} %

\put(325,-55){\thicklines\line(1,0){70}}
\put(325,-35){\thicklines\line(1,0){70}}
\put(325,-55){\thicklines\line(0,1){20}}
\put(395,-55){\thicklines\line(0,1){20}}
\put(360,-45){\makebox(0,0){\small{$\Ao=\bigcup\A_V$}}}


\end{picture}
\vspace*{30mm}

\noindent which means that to each volume $V$ it is associated an
Hilbert space $\Hil_V$ and a C*-algebra $B(\Hil_V)$, whose union
produce $\Ao$. Taking the completion of $\Ao$ wrt the C*-norm, we
get the C*-algebra of the quasi-local operators. We refer to Section
II for more details on the construction $\Hil_V$, and, in
particular, to what concerns the states on $\A$ and the dynamics.

\vspace{3mm}

The construction of the topological quasi*-algebra for a spin system
goes as follows, \cite{lass}:
\begin{enumerate}    \item let $\Hil=\mathbb{C}^2$ and, $\vec
n\in\mathbb{R}^3$, $|\vec n|=1$, and $|\vec n>\in\Hil$ fixed (but
for a phase) by requiring that $(\vec\sigma\cdot\vec n)|\vec
n>=|\vec n>$ and $|\vec n>$ is normalized. For further extensions,
it may be worth remarking that this is just a way to {\it extract}
a certain vector $|\vec n>$ out of $\Hil$.
\item let $\{\vec n_p\}$ be a sequence of normalized vectors in
$\mathbb{R}^3$ and $\{|\vec n_p>\}$ the related normalized vectors
in $\Hil_p$, all copies of $\Hil$, constructed as in 1. We put
$|\{n\}>=\otimes_{p=1}^\infty |\vec n_p>$. Of course $|\{n\}>\in
\Hil_\infty:=\otimes_{p=1}^\infty\Hil_p$, which is a non-separable
Hilbert space, \cite{thirringbook}. Also, because it is defined
via an infinite tensor product, the scalar product must be defined
with a certain care. We don't want to discuss these mathematical
details here, since they do not play a major role here, and again
we refer to \cite{thirringbook}.
\item Let $\pi$ be a natural realization of $\Ao$: $\pi(\sigma_j^\alpha)|\{n\}>=
(\otimes_{p\neq j} |\vec n_p>)\otimes (\sigma_j^\alpha|\vec
n_j>)$, and $\Hil_{\{n\}}$ be the closure in $\Hil_\infty$ of the
space $\pi(\Ao)|\{n\}>$. This is a separable Hilbert space.
\item An o.n. basis of $\Hil_{\{n\}}$ is given by the set $\{|\{m\},\{n\}>\}=\otimes_p |m_p,\vec
n_p>$, where $m_p=0,1$ for each $p$ and $\sum_p\,m_p<\infty$. Here
we have defined each vector $|m,\vec n>:=(\vec\sigma\cdot\vec
n^{\,-})^m|\vec n>$, $m=0,1$, where $\vec n^{\,-}=\frac{1}{2}(\vec
n^{\,1}-i\,\vec n^{\,2})$, $\vec n^{\,1}, \vec n^{\,2}$ and $\vec
n$ being an o.n. set in $\mathbb{R}^3$.

\item The operator $M_{\{n\}}|\{m\},\{n\}>=(1+\sum_p
m_p)|\{m\},\{n\}>$ is  unbounded, self adjoint and  greater than
$\1$. We use this to define a dense subset of $\Hil_{\{n\}}$,
$\D_{\{n\}}=D^\infty(M_{\{n\}})$, and $\D_{\{n\}}$ to define the
O*-algebra $\Lc^\dagger(\D_{\{n\}})$.

\item We find that $\pi(\Ao)\subset \Lc^\dagger(\D_{\{n\}})$.
\item We can introduce a topology $\tau_0$ on $\Ao$ as follows: $\forall
X\in\Ao$ we put, as usual, $$
\|X\|^{f,k}_{\{n\}}:=\max\left\{\|f(M_{\{n\}})\pi(X)M_{\{n\}}^k\|,\|M_{\{n\}}^k\pi(X)f(M_{\{n\}})\|\right\}.
$$
As we see, these seminorms are labeled by {  $(f,k)$ and by
$\{n\}$}.
\item Taking the completion $\A$ of $\Ao$ wrt the topology $\tau_0$ we
get a topological *-algebra. The realization of $\Ao$ can be
extended to $\A$ and we find that
$\hat\pi(\A)\subset\Lc^\dagger(\D_{\{n\}})$.
\item In the analysis of concrete models like the BCS model of superconductivity,
 \cite{lass}, it is necessary to
introduce a different topology, and some physically relevant
limits have to been searched in $\Lc(\D_{\{n\}},\D_{\{n\}}')$,
which is constructed as we have discussed in Section IV.

\end{enumerate}

\vspace{2mm}

The same construction can be repeated for other infinite discrete
quantum system obtained as infinite tensor product of finite
dimensional Hilbert spaces. For instance, if $dim(\Hil)=N$ and if
$\psi_j$, $j=0,1,\ldots,N-1$ is an o.n. basis of $\Hil$, we can
still take a vector $\Psi_0=\otimes_{p\in\mathbb{Z}}\,\psi_{0,p}$
which belongs to the non separable Hilbert space $\Hil_\infty$
constructed as before. Starting from this vector we can introduce a
separable Hilbert space as the closure in $\Hil_\infty$ of
$\pi(\Ao)\Psi_0$, $\pi$ being the natural realization of the algebra
of the matrices $N\times N-$ and, finally, a {\it number-like
operator} $\hat N_\Psi$ which is unbounded, self-adjoint and densely
defined (and play the role of $M_{\{n\}}$ above). The rest of the
construction  can be easily repeated, and a topological quasi
*-algebra associated to the physical system can be finally constructed.

\newpage

\appendix
\renewcommand{\theequation}{\Alph{section}.\arabic{equation}}


 \section{\hspace{-14.5mm} Appendix 2:  general facts in functional analysis}

This appendix is devoted to list few well known facts and results in
functional analysis which are used throughout this paper, and that
we have decided to give here to keep the paper self contained.

\begin{enumerate}

\item {\bf  Operators in a Hilbert space $\Hil$}: $A$ is
defined on a domain $D(A)$ which, if $A$ is bounded, can be taken
to be all of $\Hil$. If $A$ is unbounded (i.e. if
$\sup_{\varphi\in D(A)}\|A\varphi\|=\infty$), then $D(A)$ is a
{\em proper} subspace of $\Hil$. (e.g. $D(\hat x), D(\hat
p)\subset\Lc^2(\R)$, since $xf(x)\notin\Lc^2(\R)$ for each
$f(x)\in\Lc^2(\R)$).

\item {\bf  Closed operator}: An operator $A$ is
closed iff, for each sequence $\{\varphi_n\}\subset D(A)$ converging
to $\varphi$ and such that $A\varphi_n$ converges to $\Psi$, then
$\Psi=A\varphi$.

\item {\bf  Extension and closure of an operator}: Given two operators
$A_1$ and $A_2$ on $\Hil$ we say that $A_1$ is an {\em extension} of
$A_2$, and we write $A_1\supset A_2$,  if $D(A_1)\supset D(A_2)$ and
$A_1\varphi=A_2\varphi$ for each $\varphi\in D(A_2)$. An operator
$A$ is said to be {{\it closable}} if it has a closed extension.
Every closable operator has a smallest closed extension, called its
{\em closure}: $\overline{A}$.  $A$ is closable iff for each
sequence $\{\varphi_n\}\subset D(A)$ converging to $0$ and such that
$A\varphi_n$ converges to $\Psi$, then $\Psi=0$.

\item {\bf  Adjoint of an operator, bounded case}: in this
case $A^*$ is defined  as $<f,A^*g>=<Af,g>$, for each $f,g\in\Hil$.
If $A=A^*$ then $A$ is self-adjoint.

\item {\bf  Adjoint of an operator, unbounded case}: Again
we put $<f,A^*g>=<Af,g>$, for each $f\in D(A)$ and $g\in D(A^*)$,
where $D(A^*)=\{g\in\Hil:\,\exists g_A\in\Hil$ such that
$<f,g_A>=<Af,g>\}$. Obviously we have $g_A=:A^*g$.

\item {\bf  Symmetric operator}: let $A$ be densely defined
in $\Hil$. $A$ is {\em symmetric} if $A\subset A^*$, that is, if
$D(A)\subset D(A^*)$ and $A\varphi=A^*\varphi$ for each
$\varphi\in D(A)$. Equivalently, $A$ is symmetric if
$<Af,g>=<f,Ag>$, for each $f,g\in D(A)$. $A$ is self-adjoint if
$A$ is symmetric and if $D(A)=D(A^*)$. A symmetric operator $A$ is
called {\em essentially self-adjoint} if its closure
$\overline{A}$ is self-adjoint. In this case there exists only one
self-adjoint extension of $A$.

\item{\bf  Density matrices and traces}: A density matrix,
$\rho$, is an operator on $\Hil$ defined as $\rho=\sum_{n=1}^\infty
w_n\,P_{\psi_n}$, where $P_{\psi_n}$ are orthogonal projectors on
the o.n. set $\{\psi_n\}$ and $w_n\geq 0$ with $\sum_{n=1}^\infty
w_n=1$. Therefore $\rho$ is bounded and positive. Clearly
$tr(\rho)=\sum_{n=1}^\infty <\psi_n,\rho\psi_n>=\sum_{n=1}^\infty
w_n=1$. Remind that $tr$ does not depend on the choice of o.n.
basis.

\item {\bf  Spectral analysis}: if $A=A^*$ has a discrete
spectrum then it can be written as
$A=\sum_{n=1}^\infty\,\lambda_n\,P_{\psi_n}$, where
$\{\lambda_n\}$ and $\{\psi_n\}$ are the eigenvalues and the
eigenvectors of $A$. If $A$ has not discrete spectrum then we have
$A=\int\,\lambda\,dE(\lambda)$, in a weak sense, where
$\{E(\lambda)\}$ is a family of mutually commuting operators, such
that $E(-\infty)=0$, $E(\infty)=\I$, $E(\lambda)\leq E(\lambda')$
if $\lambda\leq\lambda'$, and $E(\lambda)\rightarrow E(\lambda')$
if $\lambda\rightarrow\lambda'$ from above. (if $A$ has discrete
spectrum then $E(\lambda)=\sum_{\lambda_n<\lambda}\,P_{\psi_n}$)

\item {\bf  One-parameter unitary groups: Stone's theorem}:
A one parameter group of unitary transformations of $\Hil$ is a
family $U_t$ of unitary operators in $\Hil$, $t\in\R$, such that
$U_tU_s=U_{t+s}$ and $U_0=\I$. This is strong-continuous if
$U_t\rightarrow\I$ strongly when $t\rightarrow0$. In this case
Stone's theorem states that there exists an unique self-adjoint
operator $K$ in $\Hil$ such that, $\forall f\in D(K)$,
$$\frac{d}{dt}U_tf=iKU_tf=iU_tKf.$$ Then we can simply  write
 $U_t=e^{iKt}$. $K$ is the {\em infinitesimal generator} of the
group.

More precisely we have the following: {\it let $U_t$ be a strongly
continuous 1-parameter group of unitary operators. The vectors
$\psi\in\Hil$ for which $\lim_{t,0}\,(-i)\,\frac{U_t-\I}{t}\,\psi$
exists form a dense set $\D$ in $\Hil$. This limit defines a
self-adjoint operator $K$ which is the infinitesimal generator of
the 1-parameter group}.

A related result is the following: {\it let $K$ be a self-adjoint
operator with spectral resolution $E_\alpha$. Then the operators
$U_t=\int_{\mathbb{R}}\,e^{it\alpha} dE_\alpha$ form a 1-parameter
group of unitary operators with $K$ as infinitesimal generator}.

\item{\bf  Tensor products}:
$\Hil_1\otimes\Hil_2=\mbox{linear span}\{f_1\otimes f_2,
f_j\in\Hil_j, j=1,2\}$, with  scalar product $<f_1\otimes
f_2,g_1\otimes g_2>=<f_1,g_1>_1+<f_2, g_2>_2$. The operators
$A_j\in B(\Hil_j)$, $j=1,2$, define a bounded operator $A_1\otimes
A_2$ on $\Hil_1\otimes\Hil_2$ as $(A_1\otimes A_2)(f_1\otimes
f_2)= A_1f_1\otimes A_2f_2$. The extension to
$\otimes_{p\in\mathbb{Z}}\,\Hil_p$ is rather subtle, and it is
discussed in details in \cite{thirringbook}.

\item {\bf Strong dual topology}:
Let $E$ be a locally convex space and $F$ its dual, i.e. the set of
the bounded linear functionals on $E$. The {\it    strong dual
topology} is a topology on $F$, $\beta(F,E)$, defined by the
following seminorms:
$$
F\ni f\rightarrow \rho_A(f):=\sup_{x\in A}|f(x)|,
$$
which are labeled by the bounded subset of $E$, $A\subset E$.

For $\D$ and $\D'$ this becomes
$$
\D'\ni z\rightarrow \rho_{\cal E}(z):=\sup_{x\in {\cal E}}|<x,z>|,
$$
where $<,>$ is the form which puts in duality $\D$ and $\D'$ and
$\cal E$ is a bounded set in $\D$.

\end{enumerate}


\end{document}